\documentstyle[12pt,aaspp4]{article}

\begin{document}

\title{THE X-RAY SPECTRAL VARIABILITY OF THE SEYFERT GALAXY NGC~3227}

\author {I.M. George \altaffilmark{1, 2},
R. Mushotzky \altaffilmark{1},
T.J. Turner \altaffilmark{1, 2}, 
T. Yaqoob\altaffilmark{1, 2}, 
A. Ptak \altaffilmark{1, 3},
K. Nandra \altaffilmark{1, 4},
H. Netzer \altaffilmark{5},
}
\altaffiltext{1}{Laboratory for High Energy Astrophysics, Code 660,
	NASA/Goddard Space Flight Center,
  	Greenbelt, MD 20771}
\altaffiltext{2}{Universities Space Research Association}
\altaffiltext{3}{Present Address: Department of Physics, 
Carnegie Mellon University, 5000 Forbes Ave, Pittsburgh, PA 15213}
\altaffiltext{4}{NAS/NRC Research Associate}
\altaffiltext{5}{School of Physics and Astronomy and the Wise Observatory,
        The Beverly and Raymond Sackler Faculty of Exact Sciences,
        Tel Aviv University, Tel Aviv 69978, Israel.}

\slugcomment{Accepted for publication in {\em The Astrophysical Journal}}

\begin{abstract}
We present the results from {\it ASCA}\ observations of NGC~3227 performed
during 1993 and 1995, along with those from a {\it ROSAT}\ observation
performed pseudo-simultaneously with the former. We find the 0.6--10~keV
continuum to be consistent with a powerlaw with a photon index $\Gamma \sim
1.6$, flatter than that typically observed in Seyfert galaxies
confirming previous results. Significant Fe $K$-shell emission is 
observed during both epochs, with an equivalent width and profile
typical of Seyfert 1 galaxies. The {\it ASCA} observations in 1993 reveal
absorption by a screen $N_{H,z}^{ion} \simeq 3\times10^{21}\ {\rm cm^{-2}}$ of
ionized material with an X-ray ionization parameter $U_X \simeq 0.01$. Both
the column and ionization-state of this material are at the low end of the
distribution of parameters observed for Seyfert~1 galaxies. Joint analysis of
the {\it ASCA}\ and {\it ROSAT} data at this epoch show an additional screen
of neutral material instrinsic to NGC~3227 with $N_{H,z}^{neu} \sim
$few$\times10^{20}\ {\rm cm^{-2}}$.

We find NGC~3227 to exhibit significant spectral variability both within and
between the observations. The most likely explanation involves
short-term variability in the continuum emission and longer-term variability
in the column density of the ionized material. Time-resolved spectroscopy and
color--color analysis indicate that the slope of the continuum
steepened by $\Delta \Gamma \simeq 0.1$ during a flare of duration $\sim
10^4$s, within the 1993 observation. However we were unable to distinguish
between a steepening of the 'primary' continuum and a change in the relative
strengths of the power-law and a putative Compton-reflection component. The
 absorbing column increased by a factor of $\sim10$ by the
1995 epoch, while the continuum is consistent with that observed in 1993. The
1995 data also show evidence that the warm absorber allows $\sim$10\% of the
nuclear emission to escape without attenuation.

We review our findings in the context of the previous results from this and
similar objects and discuss the prospects of future observations. 
\end{abstract}


\clearpage
\section{INTRODUCTION}
\label{Sec:intro}

Recent X-ray spectra obtained by {\it ASCA} have revealed the presence of 
ionized material along the cylinder--of--sight in 
a large fraction (50--75\%) of Seyfert 1 galaxies (e.g. Reynolds 1997; 
George et al 1998a, hereafter G98). To-date, this material has been 
detected primarily due to the bound-free absorption edges of 
O{\sc vii}(739~eV) and O{\sc viii}(871~eV) imprinted on the underlying 
X-ray continuum, although additional 
edges due to other ions have been detected 
in some objects.
The depths of the absorption features (and hence accuracy to which they 
can be determined) vary from object-to-object, 
with sources having implied column densities 
covering a range 
$N{\rm (O{\sc vii}+O{\sc viii})} 
	\lesssim$few$\times 10^{17}\ {\rm cm^{-2}}$
to $\gtrsim 10^{19}\ {\rm cm^{-2}}$.
Both these limits on the column density 
most likely arise only as a result of the combination of the moderate 
spectral resolution of the {\it ASCA} detectors and the 
signal--to--ratio of most of the datasets currently available.
For 
$N{\rm (O{\sc vii}+O{\sc viii})} \lesssim  7\times10^{17}\ {\rm cm^{-2}}$,
the optical depths (at the respective threshold energies) of
both O{\sc vii} and O{\sc viii} are $\lesssim 0.05$,
making the unequivocal detection of such features extremely difficult.
For 
$N{\rm (O{\sc vii}+O{\sc viii})} \gtrsim 10^{19}\ {\rm cm^{-2}}$,
bound-free absorption edges due to other ionized elements 
(most notably due to C, N, Ne $K$-shell and Fe $L$-shell transitions) 
become increasingly important and finally dominate the opacity 
if the absorbing material has 'cosmic' abundances, 
making the determination of the precise strength of the 
O{\sc vii} and O{\sc viii} edges difficult
without detailed modelling.
Nevertheless, even the current limits show that the column density of
ionized material along the cylinder--of--sight far exceeds that of neutral 
material in the majority of Seyfert 1 galaxies.
Assuming a standard abundance ratio of oxygen 
(O/H$\sim 9\times 10^{-4}$) the observed range of 
$N{\rm (O{\sc vii}+O{\sc viii})} $ implies total hydrogen 
column densities for Seyfert 1 galaxies as a class covering the range 
$10^{21} \lesssim N_{H,z}^{ion} \lesssim 10^{23}\ {\rm cm^{-2}}$.
It should be stressed that these values of 
$N_{H,z}^{ion}$ are {\it lower} limits since the column density of
oxygen with ionization states $<$O{\sc vii} and of fully stripped
oxygen ions is unknown.

As first suggested by Halpern (1984), the material responsible for these 
features is generally considered to be photoionized by the intense 
radiation field of the nucleus. 
Detailed photoionization models have been successfully 
applied to the existing X-ray data.
The location and geometry of this highly ionized material is currently 
unclear.
It is also unclear 
how the material responsible for the features observed in the 
X-ray band is related (if at all) to the 'associated absorbers' 
(primarily resonant absorption lines due to Li-like species of C and N)
commonly seen in the UV band in these objects (e.g. Crenshaw 1997) 
and some attempts 
have been made to link the two (eg. Mathur 1994; Shields \& Hamann 1997).
Since the UV absorption features with cores very close to zero intensity 
(eg. Crenshaw, Maran, Mushotzky 1998)
are seen imprinted on broad emission lines, this places 
the bulk of the UV-absorber outside the broad emission 
line region (BELR).

It is becoming increasingly clear that a single screen 
of ionized gas in photoionization equilibrium may be too simple a model.
First, the column density of the ionized gas has sometimes been seen to vary 
by large factors on relatively long timescales
(e.g. MCG-6-30-15, Fabian et al 1994; 
NGC~3783 George et al 1998b), in fact 
the first paper proposing the 
presence of absorption by ionized material in an active galaxy, 
the QSO MR2251-178, made this suggestion based on the 
variable column density (Halpern 1984).
Second, differential variability between the depths of the 
O{\sc vii} and O{\sc viii} edges has been seen in some objects
on timescales $\sim 10^4$~s
(e.g. MCG-6-30-15, Reynolds et al 1995, Otani et al 1996;
NGC~4051, Guainazzi et al 1996).
Third, a number of sources have revealed spectroscopic evidence for one 
or more screens of ionized gas (e.g. NGC~3516, Kriss et al 1996; 
NGC~3783, George et al 1998b, see also those objects with a 
'1~keV deficit' in G98).
Such observations are providing our first
insights into the location, physical conditions and 
kinematics of the material in the vicinity of the X-ray source.

The nucleus of the Sb galaxy NGC~3227 ($z=0.003$)
is heavily reddened (see Komossa \& Fink 1997, thereafter K97, 
and references therein), which is at least 
partially responsible for its classification historically as both a 
Seyfert 2 (Huchra \& Burg 1992) and 
Seyfert 1.5 (Osterbrock \& Martel 1993)
from spectroscopy of the optical emission lines.
Evidence for substantial amounts of neutral material within the 
galaxy is also provided by the detection of 
H{\sc I} (Mundell et al 1995a) and 
OH (Rickard, Bania \& Turner 1982) absorption, along
with 
CO (e.g. Rigopoulou et al 1997 and references therein)
and
H$_2$CO (tentatively; Baan, Haschick \& Uglesich 1993)
emission.
The host galaxy appears to be interacting with a nearby companion, NGC~3226
(eg. Arp 1966). 
The radio, \verb+[+O{\sc iii}\verb+]+ and H$\alpha$
emission reveal asymmetrical physical conditions on the 
opposing sides of the nucleus and a 
misalignment between the collimation of the 
radio emitting plasma and that of the photoionizing continuum
(e.g. Mundell et al 1995b; Gonz\'{a}lez Delgado \& P\'{e}rez 1997).
Spatially--resolved spectroscopy of the circumnuclear
regions have revealed evidence that the BELR (and hence perhaps the 
'nucleus' itself)
is off-set from the center of rotation by $\sim$250~pc 
(Mediavilla \& Arribas 1993; Arribas \& Mediavilla 1994).

In the X-ray band, NGC~3227 was first detected in the {\it Ariel-V}
sky-survey and has been observed subsequently 
by all major X-ray instruments 
(e.g. Malizia et al 1997 and references therein).
Of particular note are the results from the {\it EXOSAT}\ 
observations, which revealed differential variability between the 
soft and medium X-ray bands (Turner \& Pounds 1989),
and the {\it Ginga}\ observations which suggested an asymmetric profile
for the Fe $K\alpha$ emission line (Pounds et al 1989; 
George, Nandra \& Fabian 1990) and evidence for an Fe $K$-shell 
absorption edge (Nandra \& Pounds 1994).
{\it ROSAT} and {\it ASCA} observations of NGC~3227 in 1993 
have revealed 
the presence of ionized material within the cylinder--of--sight 
(e.g. Ptak et al 1994; K97; Reynolds 1997; G98). 
The values derived from $N_{H,z}^{ion}$ and $U_X$ are some of the 
lowest yet reliably measured in a Seyfert galaxy, which 
may be related to the relatively low luminosity of 
the source ($\sim 10^{42}\ {\rm erg\ s^{-1}}$ in the 0.1--10~keV 
band).
The $N_{H,z}^{ion}$ is similar to that 
necessary to give rise to the reddening observed in the 
optical and UV bands if the ionized material contains 
embedded dust with a gas--to--dust ratio and composition similar 
to that seen in our Galaxy (K97; G98).

Here, for the first time,  we present the results from the analysis of an 
{\it ASCA}\ observation of NGC~3227 performed in 1995 May. We 
also describe the results from a
re-analysis of an earlier {\it ASCA}\ observation carried out in 1993 May,
along
with those from a contemporaneous {\it ROSAT}\ observation.
The observations are described in \S\ref{Sec:obs}, and the 
preliminary data reduction and temporal analysis 
in \S\ref{Sec:screening}.
We show that NGC~3227 exhibits significant energy-dependent 
variations both within and between the two epochs.
In \S\ref{Sec:mean-spectra} we consider the time-averaged spectra 
at each epoch in order to parameterize the significant change in the
observed continuum and between 1993 and 1995, and to parameterize
the Fe $K$ emission evident during both epochs.
We present a more detailed analysis of the energy-dependent variability 
within each observation in \S\ref{Sec:timeres} using 
both an X-ray color analysis and 
time-resolved spectral analysis.
In \S\ref{Sec:Discuss} we review our findings
in the context of the previous results from this and 
similar objects and 
briefly discuss the prospects in the future, and in 
\S\ref{Sec:Conc} present our conclusions.

\section{THE OBSERVATIONS}
\label{Sec:obs}

The new {\it ASCA}\ observation of NGC~3227 reported here was carried 
out over the period 1995 May 15--16, 
we have also performed
a re-analysis of the {\it ASCA}\ observation from
1993 May 08--09, described previously in
Ptak et al (1994), Reynolds (1997), Nandra et al (1997a,b),  and G98. 
Here we utilize raw data 
from the {\tt Rev2} processing\footnote{hence raw 
data from the same processing configuration (6.4.2) is used for all the 
datasets}, along with 
new screening criteria and the latest calibration files.
These changes result in slightly different values for some parameters
compared to the results published previously.
Details on the {\it ASCA}\ satellite, its instrumentation and 
performance can be found in Makishima et al. (1996) and references therein.

The {\it ROSAT} observation of NGC~3227 reported here was performed 
over the period 1993 May 08-19 with the PSPC in the focal plane.
The results from this observation have been reported previously
by K97 and are included again here to enable 
us to compare the results to the contemporaneous {\it ASCA} data.
Our analysis differs slightly from that used by K97 
since we make use of a different analysis package ({\tt FTOOLS}) and
include the latest spatial and temporal gain corrections 
for the data. 
Details on the {\it ROSAT}\ satellite, its instrumentation and
performance can be found in Briel et al. (1994) and references therein.
The full observing log is given in Table~1.

\section{DATA SCREENING \& PRELIMINARY ANALYSIS}
\label{Sec:screening}

\subsection{Data Screening}
\label{Sec:asca_screening}
\label{Sec:ros-screening}

The unscreened {\it ASCA}\ event files containing data collected in 
{\tt FAINT} and {\tt BRIGHT} data modes were combined, as were data 
obtained during all three telemetry modes.
These data were then screened using the {\tt ascascreen}/{\tt xselect} 
script (v0.39) within the {\tt FTOOLS} package (v4.0). The 
screening criteria used were as given in Nandra et al (1997a),
with the exception that the 
elevation angle above the Earth's limb was 
$>20^{\circ}$ for the SIS0 data obtained in 1993, and 
that the CCD pixel threshold was $<$50 and $<$100 
for the 1993 and 1995 data (respectively).
The original pulse-height assignment for each event was converted to 
a pulse-invariant (PI) scale using {\tt sispi} (v1.1).
In the case of the GIS data (only) 'hard particle flares' were rejected 
using the so-called {\tt HO2} count rate, and standard 'rise-time' rejection 
criteria employed.
These criteria resulted in effective exposure times of $\sim30$~ks 
and $\sim$39~ks in each SIS and GIS (respectively) during the 
1993 observations, 
and 
$\sim34$~ks and $\sim$37~ks during the
1995 observations.

The cleaned {\it ROSAT}\ event file, produced using the standard 
screening criteria 
provided by {\tt SASS} (v7.9), was extracted from the HEASARC archive.
The latest corrections were then applied for the spatial and temporal 
variations in the gain of the PSPC (see Snowden et al 1995)
using {\tt pcsasscor} (v1.1.0), 
{\tt pctcor} (v1.1.0) and {\tt pcecor} (v1.2.0).
The effective exposure amounted to 19.5~ks spread over an 11 day 
period.

\subsection{Image Extraction}
\label{Sec:spatial-xtractcells}
\label{Sec:spatial}

Images were extracted for each instrument during each observation.
In all cases, a bright source was detected with an X-ray centroid consistent
with the optical position of NGC~3227 to within the uncertainty in the
positional accuracy of the attitude reconstruction of the 
respective satellite.

As has been previously noted by Radecke (1997), a number of serendipitous 
sources are evident within the central 20~arcmin of the 
field-of-view of the {\it ROSAT}\ PSPC.
All are relatively weak (with count rates in the 0.2--2.0~keV band
a factor $\lesssim 3$\% of that from NGC~3227). 
Only one of the serendipitous sources 
(which we tentatively identify as NGC~3226, with a count rate of
$\sim 1.5\times10^{-2}\ {\rm ct\ s^{-1}}$ in the 0.2--2.0~keV band) 
lies close enough to affect the analysis of the PSPC data from 
NGC~3227, and this has been excluded from the regions used to 
extract the source and background data.
No serendipitous sources were detected in the {\it ASCA} images.
Given the weakness of NGC~3226 and all other 
sources in the immediate vicinity, we consider it unlikely they
contaminate significantly the analysis of the  
{\it ASCA}\ data from NGC~3227.

Extraction cells were defined for the subsequent temporal and spectral 
analysis. 
In the case of the {\it ASCA}\ SIS data, a 
circular extraction cell 
of radius $\sim 3.2$~arcmin was centered on NGC~3227. 
In the case of the 1995 observations, the 
image of NGC~3227 was centered close to the standard position 
on the default CCD chip of each SIS. 
The source region lies completely within the active 
region of the chip, and from the point-spread function (psf) of the 
XRT/SIS instrument, contains $\sim84$\% of the total source counts.
However, in the 1993 observation all four chips 
were active on each SIS and NGC~3227 was centered on the default CCD chip,
but within $\sim1.8$~arcmin of the corner of the chip closest 
to the intersection of the $2\times2$ CCD array.
In principle, the data collected by all 4 chips can be combined.
However, 
given the differences between the various CCDs comprising each SIS 
detector, here we choose to analyse only the photons falling on the 
nominal, better-calibrated CCD of each SIS.
Thus the extraction cells used were circles, but excluding 
all regions beyond the active area of the nominal chip.
From the psf, we estimate these cells included 
$\sim 61$\% and $\sim 45$\% of the total source counts for 
SIS0 and SIS1 respectively.
An extraction cell was defined to provide an estimate of the 
background for each SIS detector which consisted of 
the whole of the nominal CCD chip excluding a circular region 
of $\sim 4.3$~arcmin centered on the source.

For GIS data, the source region was circular of radius
$\sim 5.2$~arcmin centered on NGC~3227. 
Given the larger field--of--view of the GIS instrument, 
such a region lies completely on the detector at both epochs.
From the psf of the XRT/GIS instrument the region
contains $\sim89$\% of the total source counts.
An annulus, centered on the source and covering $\sim 5.2$--9.8~arcmin
was used to provide an estimate of the background. 
Given their larger field--of--view, these regions 
were fully located on the active area of the GIS detectors 
during both observations. 
All fluxes and luminosities\footnote{$H_0 = 50\ {\rm km\ s^{-1}\ Mpc^{-1}}$
and $q_0 = 0.5$ assumed throughout.}
quoted below ({\it not} count rates) 
have been corrected for the fraction of the source photons falling outside 
the source extraction cells and for the contamination of 
source counts in the background extraction cells.

In the case of the {\it ROSAT} PSPC data, the source region was 
circular of radius $\sim 2.3$~arcmin and the background region an annulus 
covering radii $\sim 2.3$--4.1~arcmin, both regions centered 
on NGC~3227 
(a circle, radius $\sim 0.7$~arcmin centered on 
NGC~3226 was excluded from both). 
From the psf of the XRT/PSPC instrument the source region
contains $\sim 98$\% of the total source counts.

\subsection{Basic Temporal Analysis}
\label{Sec:temporal}

Light curves were constructed for the source and background 
regions for several different energy ranges. 
To increase the signal-to-noise ratio, the light curves from 
SIS0 and SIS1 and from GIS2 and GIS3 were combined and 
then rebinned on a variety of timescales.

\begin{figure}
\plotfiddle{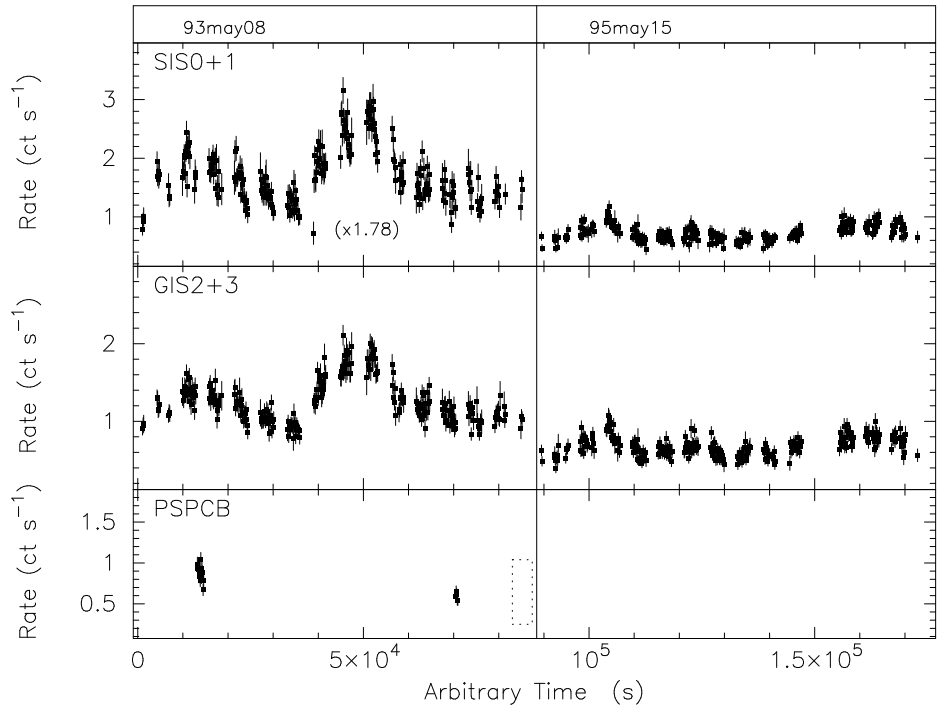}{10cm}{270}{50}{50}{-225}{300}
\caption{Light curves for the observations of NGC~3227 reported here, 
employing a bin size of 128~s. The upper two panels show the summed light
curves obtained for {\it ASCA}\ SIS and GIS detectors, whilst the lower panel
shows the portion of the {\it ROSAT}\ PSPC light curve obtained during the
1993 {\it ASCA}\ observations (there were no {\it ROSAT}\ observations
contemporaneous with the 1995 {\it ASCA}\ observations). The SIS count rate
obtained during 1993 has been rescaled in order to compensate for the smaller
extraction cell used and hence to be directly comparable to the 1995 data. The
UTC times corresponding to the first SIS data point at each epoch are 1993 May
08 03:46 and 1995 May 15 01:49. In all cases, the y-axis is scaled to cover a
factor of 0.1 to 2.5 of the mean of the 1993 light curve. Significant
variability is clearly apparent in the observed count rates both within the
individual observations, and between the 1993 and 1995 epochs. The dotted box
in the lower panel shows the total dynamic range observed in the PSPC count
rate over the period May 08--19.
\label{fig:lc128}}
\end{figure}

In Fig.~\ref{fig:lc128} we show the light curves for the 
SIS (0.5--10~keV), GIS (2--10~keV), and PSPC (0.2-2.0~keV) 
with a bin size of 128~s.
Variability is apparent in the {\it ASCA} light curves 
both within the individual observations and between the epochs.
Although only $\sim 2.3$~ks of the {\it ROSAT} pointing
were made during the 
1993 {\it ASCA} observations, variability in 
the PSPC count rate is apparent during this period 
and during the remainder of the {\it ROSAT}\ observation.
The full PSPC light curve is shown by K97, 
and the dynamic range exhibited by the PSPC count rate during the 
whole {\it ROSAT}\ observation indicated by the dotted box 
in Fig.~\ref{fig:lc128}.

\begin{figure}
\plotfiddle{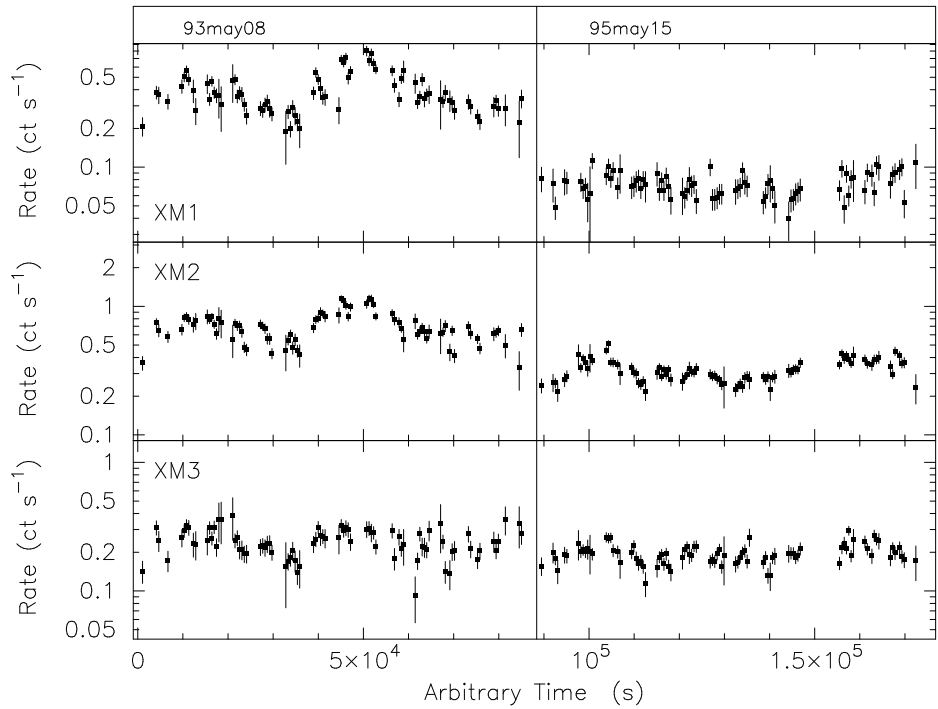}{10cm}{270}{50}{50}{-225}{300}
\caption{Light curves for NGC~3227 in 0.5--1.2~keV (XM$_1$), 1.5--3.5~keV
(XM$_2$), and 4.0-10.0~keV (XM$_3$) energy bands of the SIS employing a bin
size of 512~s. In all cases the y-axis is logarithmic,
and is scaled to cover a factor of 0.2 to 7.0 of the mean. Significant
variability is apparent both within the individual observations, and
(especially at lower energies) between the 1993 and 1995 epochs.
\label{fig:all_3227_xm512}}
\end{figure}

In Fig.~\ref{fig:all_3227_xm512} we show the light curves for the 
SIS using a bin size of 512~s using the bands
XM$_1$: 0.5--1.2~keV,
XM$_2$: 1.5--3.5~keV,
and 
XM$_3$: 4.0-10.0~keV (Netzer, Turner \& George 1994).
The count rate decreased
in all three bands between the 1993 and 1995 observations
with the amplitude increasing towards lower energies.
Clearly the source underwent a significant change in the shape of the 
observed spectrum between the two epochs.

In Table~2 we list 
the normalized 'excess variance', $\sigma^2_{rms}$, of 
each {\it ASCA} light curve using the prescription\footnote{We note there 
is a typographical error in the expression for the error
on $\sigma^2_{rms}$ given in Nandra et al (1997a) whereby 
the expression within the summation should be squared.}
given in Nandra et al (1997a).
The values of $\sigma^2_{rms}$ show 
statistically significant variability (at $>95$\% confidence)
in the XM$_1$ and XM$_2$ bands during 1993 and in the 
XM$_2$ band during 1995.
There is an indication that 
$\sigma^2_{rms}$ decreases towards higher energies
when the whole of the 1993 dataset in considered, 
but (as will be discussed in \S\ref{Sec:timeres})
$\sigma^2_{rms}$ is independent of energy when the 
'outburst' is excluded. 
In the case of the 1995 observations, $\sigma^2_{rms}$ is 
significantly lower in the XM$_1$ band
indicating the process responsible for the variability is not stationary.
We also note that $\sigma^2_{rms}$(XM$_2$) $>$ $\sigma^2_{rms}$(XM$_3$)
during the 1995 observations.

\subsection{Spectral Analysis Technique}
\label{Sec:spectral-prelims}

Source and background spectra were extracted from the cleaned event 
list of each detector using the extraction cells described in
\S\ref{Sec:spatial-xtractcells}. 
For the SIS datasets, redistribution matrices
generated using {\tt sisrmg} (v0.8) were used.
For the GIS datasets the 
redistribution matrices released on 1995 Mar 06 (generated by
{\tt gisres} v4.0) were used. 
Ancillary response files were generated for all the {\it ASCA} detectors 
using {\tt ascaarf} (v2.72).
In the case of the PSPC data, the standard response file
\verb+pspcb_gain2_256.rsp+ was used.

In all cases described below, the spectral analysis is performed on the
data from all instruments simultaneously, with different relative
normalizations to account for (small) uncertainties in the determination of
their effective areas. 
(The correction for the relative fractions of the source counts 
falling outside the source regions 
is applied within the ancillary response files.)
Data from the SIS below 0.6 keV were excluded from the spectral
analysis as it is commonly accepted that there are 
significant
uncertainties associated with the calibration of the XRT/SIS system below
this energy. 
Whilst the SIS calibration
is suspect at these energies, we do make use of the fact that 
it is considered unlikely to be in error by
$\gtrsim20$\% (see below).
The individual spectra were grouped such as to contain a minimum of 20
counts per new bin, and hence allowing $\chi^2$ minimization techniques to
be employed within the {\tt XSPEC} (v10.00) spectral analysis package.

We have adopted spectral 
models consisting of an underlying power-law continuum (of photon index
$\Gamma$) absorbed by a screen of neutral material 
at zero redshift. In all cases the column density of this material, 
identified as being due to absorption within our galaxy,
was fixed at $N_{H,0}^{gal} = 2.1\times10^{20}\ {\rm cm^{-2}}$
as derived from 21~cm measurements towards NGC~3227 (Murphy et al. 1996; these 
authors estimate an uncertainty $\lesssim 10^{19}\ {\rm cm^{-2}}$). 
All the spectral models also contain
additional screens of neutral and/or ionized material fully- or 
partially-covering the cylinder--of--sight to NGC~3227. 
With the small redshift of NGC~3227, {\it ASCA} is unable to 
distinguish absorption at $z=0$ from that at the redshift of the source.
Nevertheless, here we assume that these additional screens are 
instrinsic to NGC~3227.

\subsection{Models of the Photoionized Gas}
\label{Sec:ion}

Models including a photoionized absorber have been 
suggested previously to offer the most viable explanation for 
features observed in this and other Seyfert 1 galaxies.
The photoionization code {\tt ION} 
(version {\tt ION96}) was used to calculate the physical state of a 
slab of gas when illuminated by an ionizing continuum
(Netzer 1993, 1996). 
As in G98, below 200~eV we assume an 
illuminating continuum typical of 
AGN of the luminosity of NGC~3227 (the 'weak IR' case of Netzer 1996),
solar abundances and a density of $n = 10^{11}\ {\rm cm^{-3}}$. 
(Models with densities as low as $10^{8}\ {\rm cm^{-3}}$
give indistinguishable results.)
At energies $>$200~eV the ionizing continuum was assumed to be 
a powerlaw, and series of models were calculated assuming different 
X-ray spectral indices. 
Following Netzer (1996) and G98, the
dimensionless 'X-ray ionization parameter', $U_X$ (defined in G98, eqn. 1), 
is used to parameterize the intensity of the ionizing continuum 
in the 0.1--10~keV band.
The conversion factors between $U_X$ and ionization parameters defined over 
the entire photoionizing continuum ($>$13.6~eV) for various spectral forms 
can be found in G98.

\section{ANALYSIS OF THE TIME-AVERAGED SPECTRA}
\label{Sec:mean-spectra}

Despite the variability exhibited by NGC~3227 at both epochs
(Figs.~\ref{fig:lc128} \&~\ref{fig:all_3227_xm512}), 
it is useful to first consider the time-averaged spectra 
observed during each observation.
Spectra were therefore extracted using the extraction cells described 
in \S\ref{Sec:spatial-xtractcells} for the whole 
duration of each observation.
A strong emission line is present in the spectrum
of NGC~3227 during both observations as a result of iron $K$-shell 
fluorescence.
The best-fitting parameters of such a line are dependent
upon the form of the underlying continuum, which itself is highly
correlated with the absorption present in the soft X-ray band. 
Thus in \S\ref{Sec:mean-cont}
we exclude the 5--7~keV band (source frame) from our analysis  and
concentrate first on the form of the continuum and the nature of the absorber. 
The characteristics of the emission line and its possible effect on the 
properties of the continuum and absorption are then considered 
in \S\ref{Sec:mean-Fe}. 

\subsection{Analysis excluding the Iron $K$-shell Region} 
\label{Sec:mean-cont}

\subsubsection{The 1993 Observations}
\label{Sec:mean-1993}

\begin{figure}
\plotfiddle{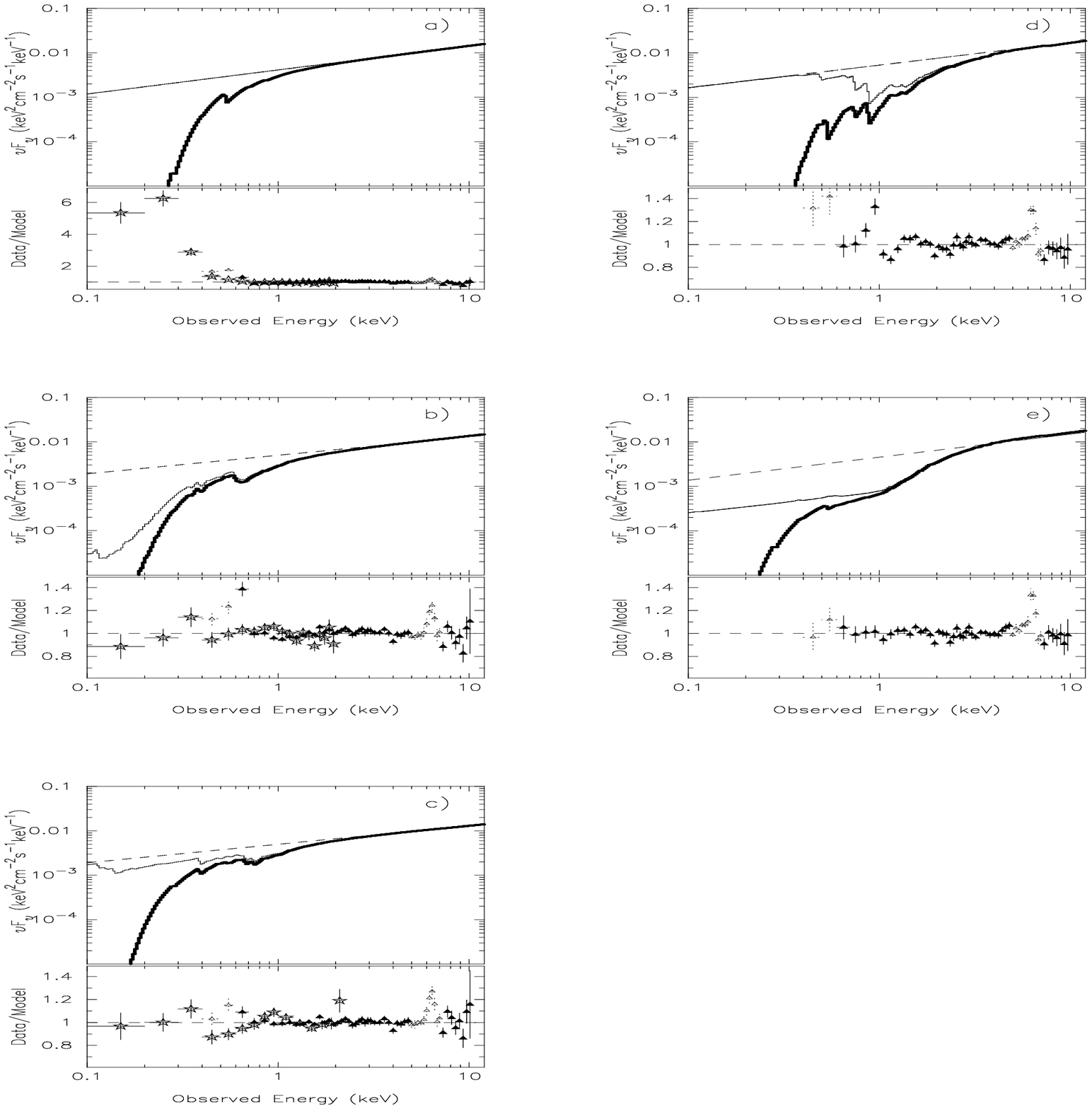}{12cm}{0}{50}{50}{10}{-200}
\caption{\it (caption on previous page)
\label{fig:mean_ufmodelratio}}
\end{figure}

We first consider a model (hereafter model~A) in which 
a single powerlaw continuum is absorbed by the Galactic column 
density $N_{H,0}^{gal}$ and an additional screen of neutral material 
($N_{H,z}^{neu}$ at $z=0.003$). 
A statistically acceptable fit for the 1993 {\it ASCA} data is obtained 
using model~A, with a $\chi^2$--statistic, $\chi^2 = 1200$ for 
1126 degrees-of-freedom (dof) and best-fitting parameters 
listed in Table~3 (Fit~1).
However, the extrapolation of this model below 0.6~keV gives rise to an 
increase in the $\chi^2$--statistic, $\Delta \chi^2_{0.6} = 43$ 
for $\Delta N_{0.6} = 12$ additional data points
(hence $\Delta \chi^2_{0.6} / \Delta N_{0.6} = 3.6$), and a mean data/model 
ratio in the 0.4--0.6~keV band of $\overline{R_{0.6}} = 2.3$.
Thus we reject\footnote{As noted in \S\ref{Sec:spectral-prelims}, the 
   calibration of the XRT/SIS system below 0.6~keV remains somewhat 
   suspect (hence our exclusion of data $<$0.6~keV during our spectral 
   analysis). However, it is unlikely that the uncertainties are of an 
   amplitude sufficient to provide a viable explanation of the values of 
   $\Delta \chi^2_{0.6} / \Delta N_{0.6}$ and $\overline{R_{0.6}}$ obtained 
   here. We adopt the criteria used in G98 whereby we consider a 
   model to extrapolate to SIS energies $<$0.6~keV in an acceptable manner if 
   {\it either} $\Delta \chi^2_{0.6} / \Delta N_{0.6} < 2.0$ 
   {\it or} $\overline{R_{0.6}}$ lies in the range 
   $0.8 \leq \overline{R_{0.6}} \leq 1.2$.
   Recent results from the cross-calibration of 
   the {\it ASCA}\ SIS and {\it BeppoSAX} LECS instrument show 
   such criteria to be reasonable (Orr et al 1998).}
model~A as an adequate description of 
the time-averaged {\it ASCA}\ spectrum during 1993.
The inadeqacy of model~A in the soft X-ray band is confirmed by a joint 
analysis of the {\it ASCA} and {\it ROSAT} data
(Table~3, Fit~2). 
From the mean data/model ratios plotted in the lower panel of
Fig.~\ref{fig:mean_ufmodelratio}a it can be seen that model~A 
underpredicts the number of counts observed $\lesssim 0.6$~keV. 

The excess of counts compared to model~A could be 
considered evidence for either a steepening of the continuum and/or an
additional emission component in the soft X-ray spectrum of NGC~3227. 
Indeed such components (with 2 or more additional free parameters) 
can be invoked to improve the quality of the extrapolation $<$0.6~keV.
However, superior fits also can be obtained with one additional 
free parameter, $U_X$, the ionization state of 
the gas instrinsic to NGC~3227. 
Hereafter we refer to this as model~B.
This model gives $\chi^2 = 1122$ for 1125 dof when applied to the 
1993 {\it ASCA}\ data alone (Table~3, Fit~4).
Not only does model~B provide 
a superior description of the {\it ASCA} data 
all energies $>$0.6~keV, it also extrapolates
in an acceptable manner to energies $<$0.6~keV 
(with $\Delta \chi^2_{0.6} / \Delta N_{0.6} = 1.7$ and
$\overline{R_{0.6}} = 0.9$).
Model~B also offers an acceptable solution to 
a joint analysis of the {\it ASCA} and {\it ROSAT} data
(Table~3, Fit~5), although it does 
appear to slightly overpredict the PSPC count rate.
(Fig.~\ref{fig:mean_ufmodelratio}b).

A vast improvement is achieved in the goodness--of-fit 
to the joint {\it ASCA}--{\it ROSAT} data 
if a screen of neutral material 
($N_{H,z}^{neu} \simeq 3.6\times10^{20}\ {\rm cm^{-1}}$)
is added to model~B
($\Delta \chi^2 = 52$).
Hereafter this is refered to as model~C. The 
best-fitting parameters of this joint {\it ASCA}--{\it ROSAT}
analysis (Table~3, Fit~8)
are consistent with those obtained from an anlysis of the 
{\it ASCA} data alone (Table~3, Fit~7), 
and are shown along with the mean data/model ratios in
Fig.~\ref{fig:mean_ufmodelratio}c.
As can be seen 
from Fig.~\ref{fig:mean_ufmodelratio}c there is a 
deficit of PSPC counts compared to such a model 
in the 0.4--0.6~keV band. No such deficit is seen in the 
{\it ASCA}\ data. 
Given the remaining uncertainties in the cross-calibration of 
the two satellites and that the source 
exhibits large-amplitude variability on short timescales (coupled 
with only a small fraction of the {\it ROSAT}\ data having been 
obtained during the {\it ASCA}\ observation), we consider 
model~C to provide a remarkably good description of the overall, 
time-averaged spectrum during the 1993 observations.
The luminosity of the underlying continuum (corrected for 
all absorption) over the 0.1--10~keV band is 
$L_{0.1-10} = (1.65\pm0.08)\times10^{42}\ {\rm erg\ s^{-1}}$, 
and that over the 0.5--2~keV band is
$L_{0.5-2} = (0.43\pm0.02)\times10^{42}\ {\rm erg\ s^{-1}}$.

\subsubsection{The 1995 Observations}
\label{Sec:mean-1995}

Neither model~A nor B provide an acceptable description of the 
the time-averaged {\it ASCA}\ spectrum obtained in 1995
(Table~3, Fits~3 \& 6).
Model~C does provide a formally
acceptable fit with $\chi^2 = 1064$ for 1001 dof
(Table~3, Fit~9), 
but does not extrapolate in an acceptable manner below 0.6~keV
($\Delta \chi^2_{0.6} / \Delta N_{0.6} = 2.1$ and
$\overline{R_{0.6}} = 1.4$) and gives rise to 
systematic residuals in the 
data/model ratios (Fig.~\ref{fig:mean_ufmodelratio}d).
Nevertheless, it is clear from the increases in 
$N_{H,z}^{neu}$, $N_{H,z}^{ion}$ and $U_X$ that 
NGC~3227 underwent a profound spectral change between 
1993 and 1995.

From further analysis of this dataset, the simplest acceptable 
model is if we allow some fraction, $D_f$, of the underlying continuum
to escape without suffering attenuation by the ionized material, whilst
the remainder ($1-D_f$) travels through an ionized column density
$N_{H,z}^{ion}$. 
We assume all the photons observed also travel through 
additional, complete screens of neutral material ($N_{H,z}^{neu}$ and 
$N_{H,0}^{gal}$).
Hereafter we refer to this model as model~D. 
Model~D provides an acceptable fit to the data
with $\chi^2 = 983$ for 1000 dof
(Table~3, Fit~12).
Furthermore this model leads to no notable systematics 
in the mean data/model ratio
and extrapolates in an acceptable manner $<$0.6~keV
(Fig.~\ref{fig:mean_ufmodelratio}e).
The current data only allow an upper limit 
($U_X \lesssim 0.06$) to be placed in the ionization parameter.
For 
the best-fitting values of $N_{H,z}^{ion}$ and $D_f$,
constraints on lower values of $U_X$ rely upon extremely 
subtle changes in the form of the observed spectrum in the 
1--3~keV band, and the data are formally consistent with $U_X = 0$
(i.e. neutral material).
The luminosity of the underlying continuum (corrected for 
{\it all} absorption) is 
$L_{0.1-10} = (1.91\pm0.10)\times10^{42}\ {\rm erg\ s^{-1}}$, 
over the 0.1--10~keV band, 
and 
$L_{0.5-2} = (0.48\pm0.03)\times10^{42}\ {\rm erg\ s^{-1}}$
over the 0.5--2~keV band.

For completeness, 
Table~3 also 
lists the best-fitting values when model~D is
applied to the {\it ASCA} data (Fit~10) and 
the joint {\it ASCA}--{\it ROSAT} data (Fit~11) obtained in 1993.
In neither case do we find a requirement for partial-covering of 
the ionized gas, although the allowed range in 
$D_f$ encompasses that found for the 1995 dataset.

\subsection{Analysis of the Iron $K$-shell Region}
\label{Sec:mean-Fe}

As can be seen from Fig.~\ref{fig:mean_ufmodelratio},
the spectrum of NGC~3227 contains an Fe $K\alpha$ fluorescence line in the 
5--7~keV band at both epochs.
We have repeated the spectral analysis (using the {\it ASCA}\ data only)
including the data within this band.
Only crude estimates of the line intensity and 
profile are possible with the current data due to the small number of 
line photons detected 
(summing all four detectors only $\sim300$ Fe $K\alpha$ photons were 
detected during the 1993 observations, and $\sim400$ photons 
during the 1995 observations). 
We assume a continuum given by model~C (\S\ref{Sec:mean-cont}) 
and have added spectral components to model the Fe emission.
In all cases we include the Fe $K\beta$ emission with the same profile as 
the $K\alpha$ component, but at 0.113 its intensity
(e.g. Kikoin 1976).

\subsubsection{Gaussian Line Profiles}
\label{Sec:mean-Fe-gauss}

First we consider simple Gaussian line profiles, with a centroid at a 
rest-frame energy $E_z(K\alpha)$ for the $K\alpha$ component,
width $\sigma$, and intensity $I(K\alpha)$.
Such a model provides an acceptable to fit to the data at both 
epochs, giving $\chi^2/dof = 1359/1399$ and $\chi^2/dof = 1205/1261$ for 
the 1993 and 1995 observations respectively.
In both cases the 90\% confidence range (for the three interesting
parameters associated with the Fe emission)
for the centroid energy is 
$6.23 \lesssim E_z(K\alpha) \lesssim 6.45$~keV
and consistent with all ionization states from Fe{\sc i}
(6.4~keV) to Fe{\sc xviii}.
The width of the line is 
constrained to lie in the range $0.0 \leq \sigma \lesssim 250$~eV
(at 90\% confidence). 
The intensity of Fe emission was similar during both epochs
($I(K\alpha) = 4.1^{+2.6}_{-1.1} \times10^{-5}\ {\rm photon\ cm^{-2}\ s^{-1}}$
and 
$5.8^{+2.1}_{-1.3} \times10^{-5}\ {\rm photon\ cm^{-2}\ s^{-1}}$
for the 1993 and 1995 observations respectively). 
As evident from 
Fig.~\ref{fig:mean_ufmodelratio}, 
the {\it observed} continuum at $\sim$6.4~keV is 
also similar 
($\sim3\times10^{-4}\ {\rm photon\ cm^{-2}\ s^{-1}\ keV^{-1}}$),
giving equivalent widths in the range 
$100 \lesssim EW(K\alpha) \lesssim 250$~eV at both epochs.
The best-fitting value of the other parameters
($\Gamma$, $N_{H,z}^{neu}$, $N_{H,z}^{ion}$, $U_X$ and $D_f$)
were consistent with the values found in 
\S\ref{Sec:mean-cont}.

We have also included a spectral component to a model any 
'Compton-reflection' of the underlying continuum. 
The reflected continuum is a flat spectral component 
(e.g. see George \& Fabian 1991). 
Given the sensitivity and bandpass of {\it ASCA}\ 
and the signal--to--noise ratio of the 
NGC~3227 data, such a component 
is unlikely to be unambiguously detected but its presence can affect 
the observed continuum $\gtrsim 6$~keV. 
We assuming this reflecting material has a 
planar geometry, is neutral, subtends a solid angle $\Omega_R$ at the 
continuum source, and is observed face-on. For simplicity we only include 
the continuum reflected by such material (any fluorescent Fe emission 
from the reflector will parameterized 
as part of the Fe emission already included in our model).
The addition of a reflection component does 
not lead to a significant improvement of the goodness--of--fit.
At both epochs
${\cal R} = {\cal F} (\Omega_R/2\pi) \lesssim 2$ (at 90\% 
confidence), 
where ${\cal F}$ is 
a scaling factor to account for any enhancement of the flux 
seen by the reflector compared to that seen by the observer.
For comparison to the 
results presented in \S\ref{Sec:timeres-xspec-reflection},
we note that 
fixing ${\cal R} = 1$, the index of the 'primary' continuum
increases (by $\Delta \Gamma \simeq 0.1$)
to $\Gamma \simeq 1.70$ and $\simeq 1.64$ during the 
1993 and 1995 observations (respectively).
The best-fitting values of the other parameters
($N_{H,z}^{neu}$, $N_{H,z}^{ion}$, $U_X$ and $D_f$) 
were consistent with the values found in 
\S\ref{Sec:mean-cont}

\subsubsection{Relativistic Line Profiles}
\label{Sec:mean-Fe-diskline}

We considered so-called 'diskline' profiles for the 
Fe emission. Following the parameterization of Fabian et al (1989),
these profiles are generated assuming a planar geometry where 
the inclination of our cylinder--of--sight with respect to 
the normal to the plane 
is given by $i$, and in which the line emissivity, $q$, 
is proportional to radius $r^{-q}$ over the range $R_i < r < R_o$, 
and zero elsewhere. 
Kinematic and general relativistic effects are included assuming 
the emitting material is in Keplarian motion around 
a Schwarzschild black hole.
As is common with {\it ASCA}\ data 
with this signal--to-noise ratio, 
we cannot constrain all the parameters of the diskline component 
simultaneously so we have 
fixed $R_i$ at the innermost stable orbit of a Schwarzschild black hole
(ie. $R_i=6r_g$, where $r_g = GM/c^2$ is the gravitational
radius of a black hole of mass $M$) and 
$R_o=10^3 r_g$.
We have also restricted $q$ to lie in the range 
$0 \leq q \leq 3$ appropriate if the Fe emission is the result of 
X-ray illumination of the disk.

A diskline profile
offers a superior description of the data 
(at $> 99$\% confidence for the 1995 data using the $F$-test) 
compared to the Gaussian 
profiles discussed in \S\ref{Sec:mean-Fe-gauss}, with
$\chi^2/dof = 1355/1398$ and $1193/1259$ for 
the 1993 and 1995 observations respectively 
At both epochs the best-fitting value of $E_z(K\alpha) \simeq 6.5$~keV, 
but with 90\% confidence limits (for the four interesting
parameters associated with the Fe emission) covering the range
6.4~keV (corresponding to Fe{\sc i}--{\sc xii}) to 
$\sim$6.6~keV (corresponding to Fe{\sc xxiv}).
The index of the emissivity, $q$, could not be well constrained by
the 1993 observations ($0 \lesssim q \lesssim 2.4$), but was better 
determined by the 1995 observations to lie in the range 
$1.9 \lesssim q \lesssim 2.4$).
In both cases, the inclination was found to be 
$i \lesssim 30$ degrees at 90\% confidence.
The intensity of Fe emission assuming such a profile is
$I(K\alpha) = 4.7^{+2.1}_{-1.8} \times10^{-5}\ {\rm photon\ cm^{-2}\ s^{-1}}$
($100\lesssim EW(K\alpha) \lesssim 230$~eV).
and 
$9.0^{+3.1}_{-2.9} \times10^{-5}\ {\rm photon\ cm^{-2}\ s^{-1}}$
($200\lesssim EW(K\alpha) \lesssim 400$~eV).
for the 1993 and 1995 observations respectively.

Again we find no
significant improvement of the goodness--of--fit 
at either epoch, when a Compton-reflection component is included in the
model, with ${\cal R} \lesssim 1.2$ 
(at 90\% confidence) during both epochs.
Again for comparison of the results presented 
in \S\ref{Sec:timeres-xspec-reflection}, we
find that fixing ${\cal R} = 1$, the index of the 'primary' continuum
increases to $\Gamma \simeq 1.70$ at both epochs, whilst 
the best-fitting values of the other parameters
($N_{H,z}^{neu}$, $N_{H,z}^{ion}$, $U_X$ and $D_f$) 
were consistent with the values found in 
\S\ref{Sec:mean-cont}

\section{TIME-RESOLVED ANALYSIS}
\label{Sec:timeres}

In \S\ref{Sec:mean-spectra} we found that adequate descriptions 
of the time-averaged spectra could be obtained at both epochs.
Here, given the rapid variability of the source 
(Figs.~\ref{fig:lc128} \&~\ref{fig:all_3227_xm512})
we investigate the variations within each observation in order to 
explore the origin of the variations.
First, in \S\ref{Sec:timeres-colors} we present an analysis of the 
X-ray colors, and in \S\ref{Sec:timeres-xspec} spectral analysis
of selected temporal ranges.

\subsection{X-ray Color Analysis}
\label{Sec:timeres-colors}

\begin{figure}
\plotfiddle{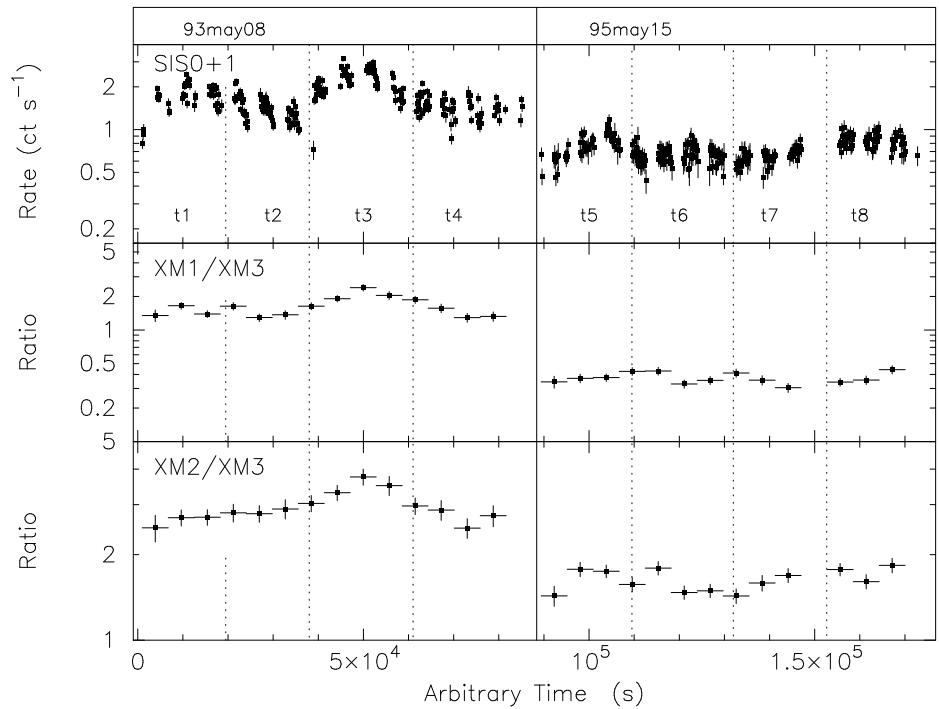}{10cm}{270}{50}{50}{-225}{300}
\caption{The Upper shows the total SIS count rate as a function of time using
128~s bins (from Fig.~\ref{fig:lc128}). The lower two panels shows the
XM$_1$/XM$_3$ and XM$_2$/XM$_3$ X-ray colors (see \S\ref{Sec:temporal})
as a function of time using
5760~s bins. In all cases the y-axis is logarithmic.
Significant variability is evident in the XM$_1$/XM$_3$ and XM$_2$/XM$_3$
colors during both epochs. During the 1993 observations it is clear that the
spectrum becomes softer as the source brightens during temporal period 1993(t3)
\label{fig:all_3227_xm5760ratio}}
\end{figure}

In \S\ref{Sec:temporal} it was found that the amplitude of the 
variability exhibited within the three spectral bands XM$_1$, XM$_2$ and 
XM$_3$ was different both within and between the 1993 and 1995 
observations (Table~2).
In Fig.~\ref{fig:all_3227_xm5760ratio} we show the 
ratio of the XM$_1$ to XM$_3$ and 
XM$_2$ to XM$_3$ count rates (hereafter referred to as the 
XM$_1$/XM$_3$ and XM$_2$/XM$_3$ colors) 
as a function of time, along with the 
total SIS count rate (from Fig~\ref{fig:lc128}) for reference.
As expected from Fig.~\ref{fig:all_3227_xm512}, 
both the XM$_1$/XM$_3$ and
XM$_2$/XM$_3$ colors are variable during the 1993 observations 
(at $>99$\% confidence). 
This variability appears correlated with 
the total SIS count rate in the sense that the source is 
softer when brighter (Fig.~\ref{fig:all_3227_xm5760ratio}).
Significant variability is also exhibited in the 
XM$_1$/XM$_3$ (at $\sim 95$\% confidence) and 
XM$_2$/XM$_3$ (at $\sim 99$\% confidence) bands during the 1995
observations, though clearly of a lower amplitude than
those seen in 1993.
Given the smaller amplitude of the variation
in the total SIS count rate during 1995, it is unclear whether 
any correlation between the color and intensity exists during 
this epoch.

\begin{figure}
\plotfiddle{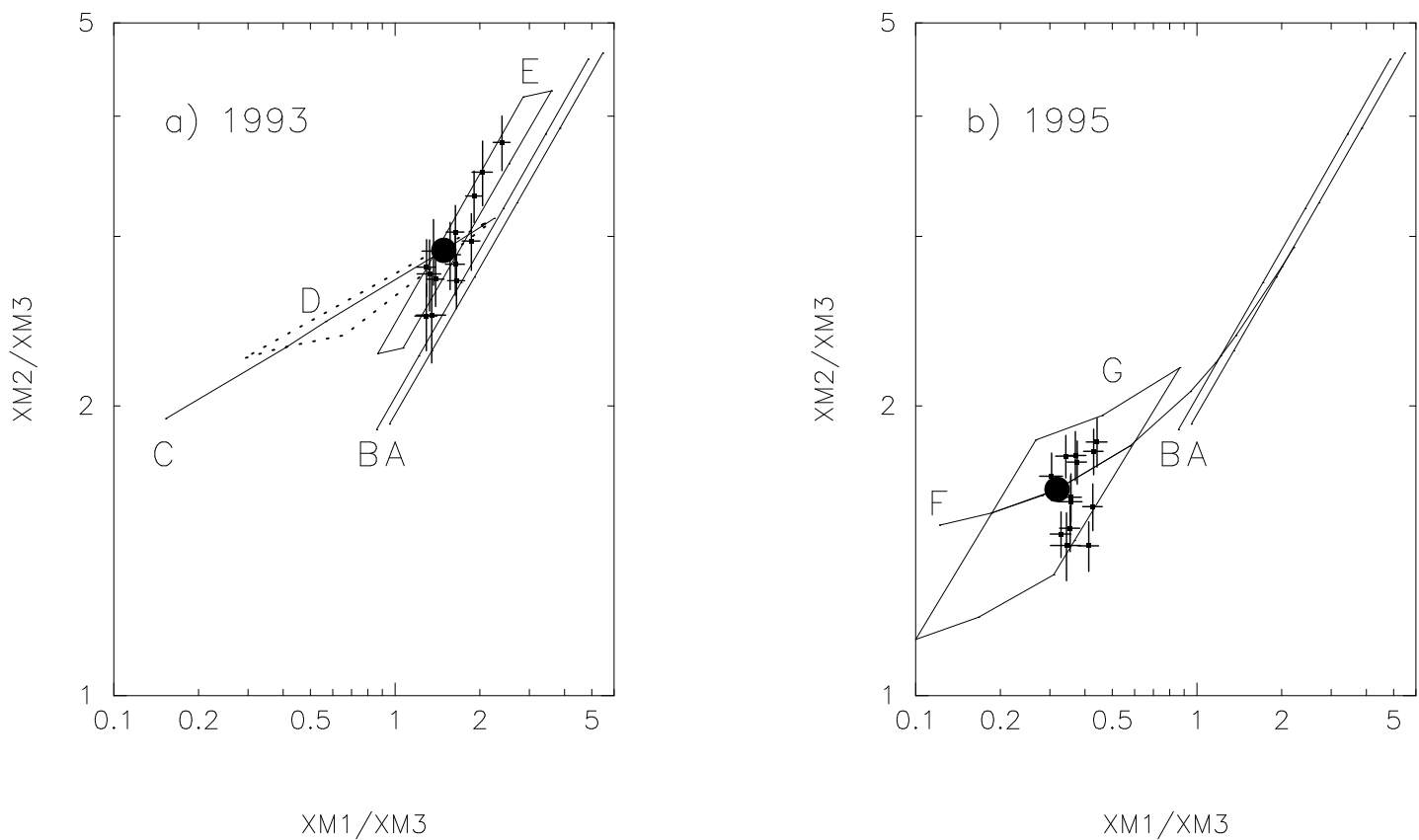}{10cm}{0}{50}{50}{10}{-200}
\caption{The XM$_1$/XM$_3$ -- XM$_2$/XM$_3$ color-color diagram for the epochs
of the NGC~3227 observations. Also shown are the predictions from various
theoretical spectra after being folded through the spectral response of the
XRT/SIS instrument (described in \S\ref{Sec:timeres-colors}). 
The filled circle 
indicates the location of the best-fitting model for the 
time-averaged spectrum described in \S\ref{Sec:mean-spectra}.
During the 1993 observations (left panel), the points with the
highest values of both XM$_1$/XM$_3$ and XM$_2$/XM$_3$ occur during the
temporal period 1993(t3) defined in Fig.~\ref{fig:all_3227_xm5760ratio}.
\label{fig:xmallover3}}
\end{figure}

In Fig.~\ref{fig:xmallover3} we show these data on the 
XM$_1$/XM$_3$--XM$_2$/XM$_3$ plane. Also 
shown are the predictions for various theoretical 
spectra after being folded through the spectral response 
of the XRT/SIS instrument.
As suggested by Netzer et al (1994),
color--color diagrams of this type can be useful in comparing the 
color variations observed with those predicted assuming different scenarios.
In both panels of Fig.~\ref{fig:xmallover3}, the filled circle 
indicates the location of the best-fitting model for the 
time-averaged spectrum described in \S\ref{Sec:mean-spectra}.
The two straight, solid lines show the loci of simple power laws
absorbed only by a screen of neutral material.
Line A is for the case where the power law is attenuated by
a column density of 
$N_{H,0}^{gal} = 2.1\times10^{20}\ {\rm cm^{-2}}$, 
whilst line B shows the effect where an additional screen of neutral 
material with a column density 
$N_{H,z}^{neu} = 3.6\times10^{20}\ {\rm cm^{-2}}$
(as found from the {\it ASCA}/{\it ROSAT} joint analysis during 
1993) also attenuates the spectrum.
Both lines are shown from $\Gamma = 1.0$ (highest 
value of XM$_1$/XM$_3$) to $\Gamma = 2.0$ (lowest value of XM$_1$/XM$_3$).

\subsubsection{The 1993 observations}

The simplest scenario is that the spectral variability 
arises solely from changes in the ionization state of a single 
screen in response to changes in continuum flux. However, 
even large changes in flux will give only small changes 
in the XM$_1$ count rate  for 
a column $N_{H,z}^{ion} \sim 10^{21} \ {\rm cm^{-2}}$  
of ionized material in equilibrium (e.g. see Netzer et al 1994).
For such a column (and $\Gamma \sim 1.6$) variations in $U_X$ from 
$10^{-3}$ to $10$
(i.e. a factor of $10^4$ in the intensity of the 
illuminating continuum) only change 
XM$_1$/XM$_3$ from $\sim 1.3$ to $\sim 2.4$
and 
XM$_2$/XM$_3$ from $\sim 2.9$ to $\sim 3.2$.
Of course detailed 
predictions of the behaviour of the gas depend on whether it has 
reached ionization equilibrium following a change in continuum intensity. 
The timescales required to reach equilibrium will be further discussed in 
\S\ref{Sec:equilibrium}. For simplicity, in the following we assume 
the ionized material is in equilibrium.

Large changes in the XM$_1$/XM$_3$ and XM$_2$/XM$_3$ colors 
can be obtained through variations in $N_{H,z}^{ion}$. 
In Fig.~\ref{fig:xmallover3}a), line C shows the locus 
of variations in  $N_{H,z}^{ion}$ (only) from 
$N_{H,z}^{ion} = 0$ (where line C intersects line B) to 
$N_{H,z}^{ion} \sim 20\times10^{21}\ {\rm cm^{-2}}$.
The dotted region marked D delineates the area on the color--color diagram 
in which both $N_{H,z}^{ion}$ and $U_X$ are allowed to vary by a 
factor 0.25--4 from the mean value.
Clearly variations of reasonable amplitude in 
neither $N_{H,z}^{ion}$ alone, nor $N_{H,z}^{ion}$ 
and $U_X$ can be the sole explanation of the observed
variability.

The parallelogram marked E in Fig.~\ref{fig:xmallover3}a)
delineates the region  on the color--color diagram covering 
a factor 0.25--4 from the mean value of
$U_X$ and spectral index in the range $1.3 \leq \Gamma \leq 2.0$,
with $N_{H,z}^{ion}$ fixed at the mean value.
The observed colors are in much better 
agreement with this region. Thus we suggest that the 
dominant cause of the variations in the 
XM$_1$/XM$_3$ and XM$_2$/XM$_3$ colors observed during 1993 
is variations in the underlying spectral index.
As evident from Fig.~\ref{fig:all_3227_xm5760ratio},
the XM$_1$/XM$_3$ and XM$_2$/XM$_3$ colors are both highest 
during the temporal range 1993(t3) suggesting the largest variation
in the underlying spectral index occurred during this period. 
Indeed as found in \S\ref{Sec:temporal}, the excess variance
($\sigma^2_{rms}$) is a strong function of energy when all the 
data from this epoch are considered, with 
$\sigma^2_{rms}$(XM$_1$)/$\sigma^2_{rms}$(XM$_3$)
$\simeq 5.6\pm2.7$ (Table~2).
When the temporal range 1993(t3) 
is excluded from the analysis, however, 
$\sigma^2_{rms}$ is consistent with being independent of 
energy 
($\sigma^2_{rms}$(XM$_1$)/$\sigma^2_{rms}$(XM$_3$)
$\simeq 1.0\pm0.7$.
The behavior of the source is therefore consistent with a steepening 
of the continuum during the 'outburst' during 1993(t3). 
This is discussed further in \S\ref{Sec:timeres-xspec}.

\subsubsection{The 1995 observations}

For the parameter-space indicated by the best-fitting 
model, the ionized absorber only has a significant effect on the 
XM$_2$ band (the XM$_3$ band is dominated by the transmitted continuum,
whilst the XM$_1$ band is dominated by the unattenuated continuum, see 
Fig.~\ref{fig:mean_ufmodelratio}d. 
Only changes in $N_{H,z}^{ion}$ by an 
order of magnitude 
will have a significant effect on the observed X-ray colors.
Similarly, the X-ray colors are insensitive to $U_X$ for a 
constant absorbing column density of 
$N_{H,z}^{ion} \sim 3\times10^{21}\ {\rm cm^{-2}}$.
On the other hand, the X-ray colors are very sensitive to 
the spectral index of the underlying continuum ($\Gamma$) and 
the fraction of this continuum allowed to escape without 
attenuation ($D_f$).

In Fig.~\ref{fig:xmallover3}b), curve F shows the locus of 
variations in  $D_f$ (only) from 
$D_f = 1.0$  (where curve F intersects line A)
to $D_f = 0.03$.
The region G delineates the area on the color--color diagram covering
a factor 0.5--2 from the mean value of
$D_f$ and spectral index in the range $1.1 \leq \Gamma \leq 1.7$,
with $N_{H,z}^{ion}$ and $U_X$ fixed at their mean values.
The observed colors are in good 
agreement with this region. Thus we suggest that the 
major cause of the variations in the 
XM$_1$/XM$_3$ and XM$_2$/XM$_3$ colors observed during 1995 
is, again, variations in the underlying spectral index, possibly accompanied 
by relatively small variations in $D_f$.

\subsection{Spectral Analysis}
\label{Sec:timeres-xspec}

Given the evidence for variability in the slope of the underlying 
continuum found from the color--color analysis in 
\S\ref{Sec:timeres-colors}, here we report the results from the
spectral analysis of selected temporal ranges.
This allows us to test the results from the color--color analysis 
and investigate a number of implications.
The temporal ranges are denoted by 1993(t1)--(t4), 1995(t5)--(t8), 
and are shown in Fig.~\ref{fig:all_3227_xm5760ratio}.
As in \S\ref{Sec:mean-cont}, we initially exclude the 5--7~keV band from this 
analysis to avoid the effects of the Fe $K$-shell emission.

\begin{figure}
\plotfiddle{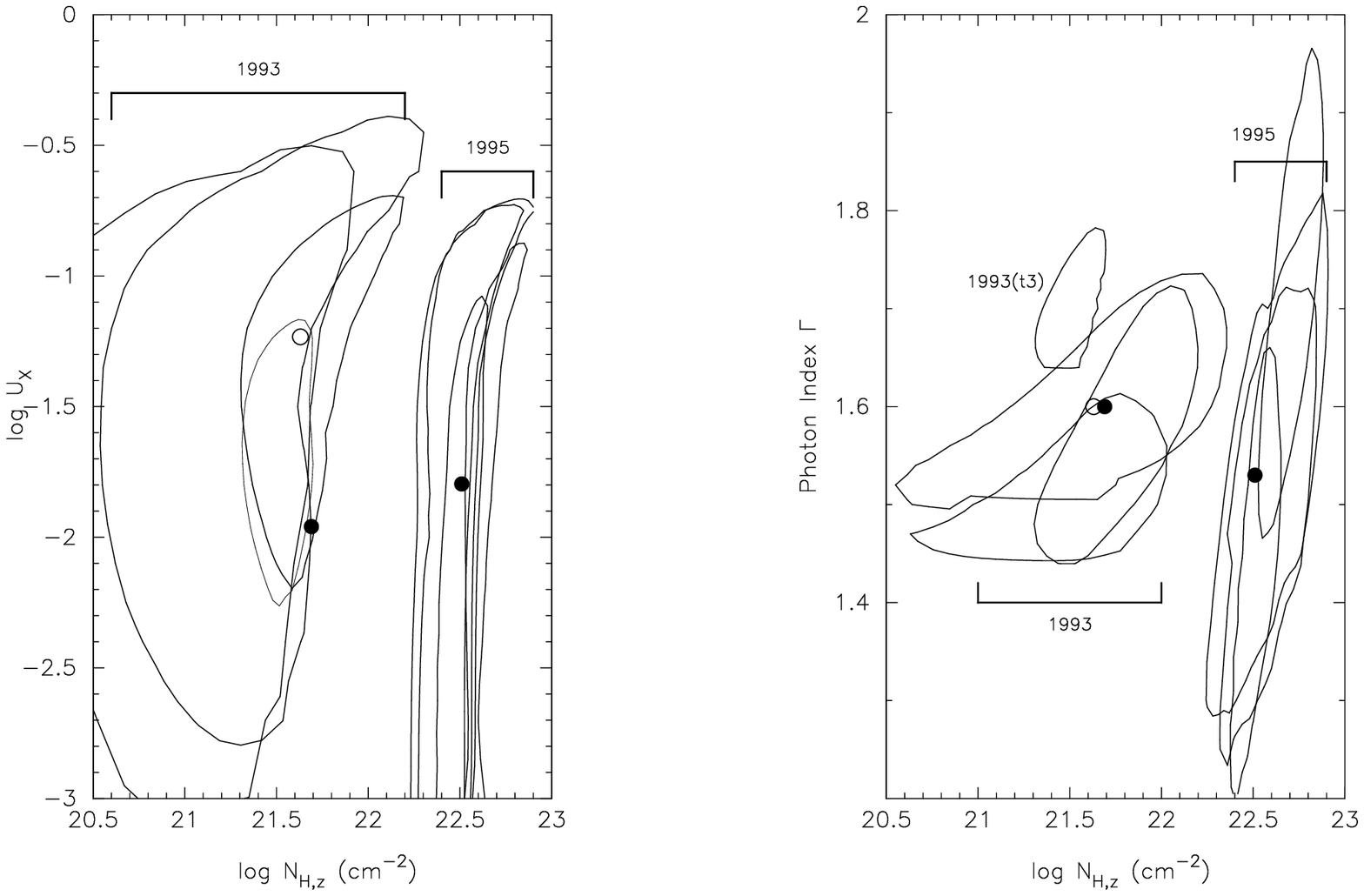}{12cm}{0}{50}{50}{-175}{0}
\caption{Contours showing the 90\% confidence regions in the $N_{H,z}$--$U_X$
and $N_{H,z}$--$\Gamma$ planes from the time-resolved spectral analysis of
NGC~3227 described in \S\ref{Sec:timeres-xspec}. The filled circles indicate
the best-fitting values from the analysis of the time-average spectra
described in \S\ref{Sec:mean-cont}. During the
temporal period 1993(t3) there underlying powerlaw appears to be steeper (by
$\Delta \Gamma \sim 0.1$) than during the remainder of the observations at
that epoch. The open circles indicate the best-fitting values when the
1993(t3) data is excluded from the analysis (see \S\ref{Sec:timeres-xspec}).
\label{fig:contours}}
\end{figure}

In Table~4 we list the results assuming our best-fitting
generalized model found in \S\ref{Sec:mean-cont}. This model consists of an
underlying powerlaw absorbed by two screens of neutral material
($N_{H,0}^{Gal}$ and $N_{H,z}^{neu}$) and a screen of ionized material
($N_{H,z}^{ion}$ and $U_X$), with a fraction $D_f$ of the continuum escaping
within suffering attenuation by $N_{H,z}^{ion}$. This 
model provides an acceptable description of all the time-resolved spectral
datasets, and extrapolates in an acceptable manner $<$0.6~keV. The contours of
the 90\% confidence regions projected onto the $N_{H,z}^{ion}$--$U_X$ and
$N_{H,z}^{ion}$--$\Gamma$ planes are shown in Fig~\ref{fig:contours}. Despite
the large uncertainties arising as a result of the reduced signal--to--noise
ratio in the time-resolved spectra, the two epochs are clearly distinct in
this three-dimensional model space. From both Table~4
and Fig~\ref{fig:contours} it can be seen that this separation is primarily
the result of an increase in column density of the absorbing material by
approximately an order of magnitude (from 
$N_{H,z}^{ion}\sim 3\times10^{21}\ {\rm cm^{-2}}$ to 
$\sim 30\times10^{21}\ {\rm cm^{-2}}$) between the the
1993 and 1995 observations. From the $N_{H,z}^{ion}$--$U_X$ plane
(Fig~\ref{fig:contours}a) it can be seen that the datasets for both epochs are
consistent with the values derived from the analysis of their respective
time-averaged spectra (filled circles), and all are consistent 
with an ionization parameter in the range $0.006 \lesssim U_X \lesssim 0.06$. 
However, on the projection onto the
$N_{H,z}^{ion}$--$\Gamma$ plane (Fig~\ref{fig:contours}b), the 1993(t3)
dataset is inconsistent with two of the 
other datasets obtained during that epoch. 
During the temporal period 1993(t3) the observed spectral index is 
steeper ($\Delta \Gamma \sim$0.1--0.2) than that derived from the analysis 
of the time-averaged spectrum during this epoch.

\subsubsection{The Effects of Excluding the 1993(t3) Data}
\label{Sec:timeres-xspec-reflection}

\begin{figure}
\plotfiddle{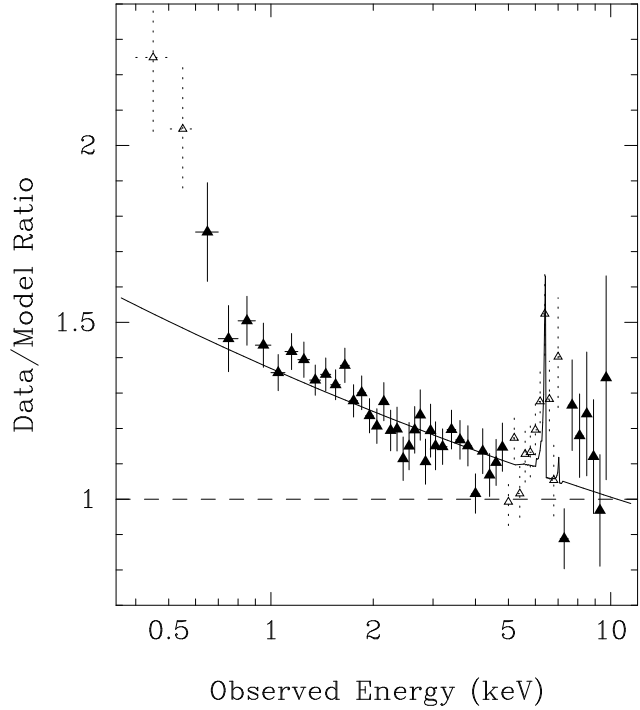}{10cm}{0}{50}{50}{-175}{0}
\caption{The mean ratio of the 1993(t3) data to the best-fitting model to the
combined 1993(t1,t2,t4) datasets ($N_{H,z}^{neu} = 1.1\times10^{21}\ {\rm
cm^{-2}}$, $\Gamma = 1.60$, $N_{H,z}^{ion} = 4.3\times10^{21}\ {\rm cm^{-2}}$,
$U_X = 0.058$, and $D_f = 0$). The solid curve shows the theoretical ratio
assuming a steepening of the underlying continuum by $\Delta \Gamma = 0.13$
pivoting at 10~keV, plus that due to the best-fitting diskline profile ($E_{z}
= 6.43$~keV, $\beta = 1.98$, $i=0$, and equivalent width 170~eV). It can be
seen that the the 1993(t3) data are consistent with such an hypothesis.
However as discussed in \S\ref{Sec:timeres-xspec-reflection}, this steepening
may also be the result of a lack of response in a Compton-reflector during the
1993(t3) 'flare'
\label{fig:t3_over_rest_ratrebin}}
\end{figure}

We have repeated the spectral analysis of the {\it ASCA}\ data
obtained during 1993,
but excluding the data obtained during the temporal period 1993(t3).
We find an acceptable fit ($\chi^2/dof = 999/998$) with
an underlying spectral index 
$\Gamma = 1.60^{+0.08}_{-0.05}$, 
absorption due to neutral material of column density
$N_{H,z}^{neu} = (1.1^{+0.7}_{-1.1})\times10^{21}\ {\rm cm^{-2}}$,
plus absorption by ionized material with parameters 
($N_{H,z}^{ion} = 
(4.3^{+3.7}_{-0.9})\times10^{21}\ {\rm cm^{-2}}$, 
$U_X = 0.05^{+0.06}_{-0.06}$)
consistent with those 
obtained from our analysis of the 
time-averaged spectrum at this epoch (\S\ref{Sec:mean-1993}).
The constraints on any unattenuated continuum provided by these data is
$D_f \lesssim 30$\% (at 90\% confidence).
The location of the best-fitting model on the 
$N_{H,z}^{ion}$--$U_X$ and 
$N_{H,z}^{ion}$--$\Gamma$ planes are shown by the open circles in 
Fig~\ref{fig:contours}.
In Fig.~\ref{fig:t3_over_rest_ratrebin} we compare 
the 1993(t3) data to this model.
The spectrum obtained during the 
1993(t3) brightening is indeed consistent with 
a steepening in the {\it observed} spectral index, 
pivoting around 10~keV.
Returning the 5--7~keV band to the analysis and assuming a 
diskline profile for the Fe emission (\S\ref{Sec:mean-Fe-diskline}), 
we find an acceptable fit ($\chi^2/dof = 1190/1204$) when the intensity 
of the $K\alpha$ line 
$I(K\alpha) = (5.1^{+2.4}_{-1.8})\times10^{-5}\ {\rm photon\ cm^{-2}\ s^{-1}}$
($EW(K\alpha) = 170^{+80}_{-60}$~eV) and
$\beta = 2.0^{+0.5}_{-0.7}$.
The upper limits on the energy and inclination of the putative disk are 
$E_{z} \lesssim 6.59$~keV
and
$i \lesssim 45$~degrees (at 90\% confidence, respectively).
The best-fitting values of the other parameters
($N_{H,z}^{neu}$, $N_{H,z}^{ion}$, $U_X$ and $D_f$) 
were consistent with the values found above.

The detection of a steepening of the spectral index of the underlying continuum 
in an AGN, on a timescale $\sim 10^4$~s and a pivot at $\sim$10~keV, may be 
highly pertinent to our understanding of the mechanism giving rise to the
'primary' X-ray continuum (\S\ref{Sec:disc-cont}). Since such a steepening is
inferred from a comparison between the data obtained during 1993(t3) and those
obtained during the remaining temporal periods of the 1993 observations, it is
important to investigate whether we might be being misled. 
Given the bandpass of {\it ASCA}\ and the signal--to--noise ratio of the 
1993 data, the delusory effect most likely is Compton Reflection of the
the primary continuum.
As stated above, the lack of effective area $\gtrsim 10$~keV
makes the amplitude of such a component difficult to constrain with 
{\it ASCA}\ with this signal--to--noise ratio.
We find that the inclusion of a Compton-reflected
continuum (with the inclination of reflector equal to
that in the Fe $K\alpha$ emission line) 
offers no significant improvement to the 'quiescent'
spectrum observed during 1993 (i.e. excluding 1993(t3)), with an
upper limit of ${\cal R} \sim  1.6$ (at 90\% confidence).
Fixing ${\cal R} = 1$, however, the index of the
underlying powerlaw to steepen to $\Gamma \simeq 1.71$ 
(and the best-fitting value of all other parameters
consistent with those found when ${\cal R} = 0$).
A similar spectral index was found from a similar analysis of 
the time-averaged spectra at both epochs in 
\S\ref{Sec:mean-Fe}.
More significantly however, with ${\cal R} = 1$, the spectral 
index is consistent with that obtained for the 
1993(t3) dataset alone ($\Gamma \simeq 1.73$, Table~4).
We conclude that we are unable to make any definitive statements regarding 
variations in the 'primary' X-ray continuum with these data
due to uncertainties in the amplitude of any Compton-reflection.

\section{DISCUSSION}
\label{Sec:Discuss}

The {\it ASCA}\ observations performed in 1993 and 1995 
reveal NGC~3227 to have exhibited significant spectral variability 
both within the 1993 observation and between the two epochs.
Detailed analysis of the data obtained during 'quiescent' periods 
of the 1993 observation
show the spectrum can be adequately described by a powerlaw 
continuum (with $\Gamma \sim$1.6). Imprinted 
on this continuum are absorption features due to ionized 
gas (with 
an equivalent hydrogen column density 
$N_{H,z}^{neu} \simeq 3\times10^{21}\ {\rm cm^{-2}}$
and X-ray ionization parameter $U_X \simeq 0.01$) and 
Fe $K$-shell emission (with $EW(K\alpha) \simeq 170$~eV).
The observed continuum steepened (to $\Gamma \simeq 1.7$) 
during the 'flare' comprising 
the temporal period 1993(t3). Unfortunately the data do not allow 
us to determine whether this represents a true steepening of 
the underlying continuum, or whether the 
index observed during 
the quiescent periods is flatter due to the presence of 
a Compton-reflector.
From a joint analysis of the {\it ASCA}\ and {\it ROSAT} PSPC 
data at this epoch we also find evidence for 
an additional screen of neutral material
with $N_{H,z}^{neu} \simeq 4\times10^{20}\ {\rm cm^{-2}}$.
The spectrum at energies $\gtrsim 3$~keV observed during the 
1995 observations is identical to that observed in 
1993. 
The spectrum at lower energies is, however, dramatically 
different and consistent with $\simeq 90$\% of the underlying 
continuum suffering attenuation by material with 
$N_{H,z}^{ion} \simeq 3\times10^{22}\ {\rm cm^{-2}}$
whilst the remaining fraction of the continuum 
is absorbed by neutral material with 
$N_{H,z}^{neu} \lesssim 10^{21}\ {\rm cm^{-2}}$.

In this section we review these findings 
in more detail, compare them to previous results 
and briefly discuss the prospects in the future.

\subsection{The Underlying Continuum}
\label{Sec:disc-cont}

The slope of the {\it observed} continuum in NGC~3227
($\Gamma \sim$1.5--1.7) is flatter than that seen in the majority of 
Seyfert galaxies ($\Gamma \sim 1.9$; Nandra \& Pounds 1994), consistent 
with the findings of most previous measurements of this source 
(Turner \& Pounds 1989; Turner et al 1991; Weaver, Arnaud \& Mushotzky 1995). 
Furthermore, during the temporal period 1993(t3) we found clear evidence 
that the observed continuum steepened (by $\Delta \Gamma \sim 0.1$, 
on a timescale $\sim$few$\times 10^4$~s) as the source brightened
(by $\Delta L_{0.1-10} \sim 40$\%).
If this steepening in the observed continuum is in fact due to a steepening 
in the 'primary' X-ray continuum, then it is of relevance to our understanding 
of the generation of the X-ray continuum.
The most popular model for the production of the underlying continuum in 
the X-ray band in Seyfert galaxies is that it 
produced by the Compton upscattering of lower energy photons. The electrons 
(and possibly positrons) responsible for this upscattering could be within
either a highly relativistic, nonthermal plasma (e.g. Svensson 1994)
or mildly relativistic, thermal plasma (e.g. Sunyaev \& Titarchuck 1980;
Ghisellini \& Haardt 1994).
Given {\it CGRO}\ observations of a break in the spectra of 
some Seyfert galaxies in the $\sim$50--300~keV
regime (e.g. Johnson et al 1993), attention has shifted recently back to 
thermal (or quasi-thermal) models (Svensson 1996 and references therein).
Several geometries have been proposed for the UV and Compton scattering 
regions, with a popular configuration having an 
accretion disk emit the UV seed photons, and a fraction of these 
photons are up-scattered into the X-ray band in a (possibly patchy) corona 
above the disk
(e.g. Stern et al 1995; Haardt, Maraschi, Ghisellini 1997).
Interestingly, as discussed by Haardt, Maraschi, Ghisellini (1997), 
a pivoting of the spectrum at $\sim 10$~keV accompanying 
relatively small changes in the X-ray luminosity 
(factor $<$2) is predicted by models in which 
the optical depth to $e^{\pm}$-pair production is 
$\lesssim 0.1$ and weakly coupled to the coronal luminosity.

Whilst we find no statistical requirement for a Compton-reflection component
in NGC~3227, the possible presence of such a component cannot be excluded,
with ${\cal R} \lesssim 1.2$ at 90\% confidence (\S\ref{Sec:mean-Fe-diskline}).
Previous {\it Ginga}\ observations find ${\cal R} \simeq 1\pm0.5$ 
(Nandra \& Pounds 1994), and the strength of the Fe $K$-shell emission 
($EW(K\alpha) \sim$100--250~eV, \S\ref{Sec:mean-Fe}) provides circumstantial 
evidence for ${\cal R} \gtrsim$0.5 (e.g. George \& Fabian 1991).
Preliminary results from a recent 
{\it RXTE}\ observation suggest a relatively weak Compton-reflection 
component in NGC~3227 (Ptak et al 1998).

Finally we note that whatever the cause of the spectral
variability observed during 1993(t3), 
when these data are excluded from 
the analysis the source still exhibits relatively large 
values of the excess variance $\sigma^2_{rms}\sim 0.03$
over the full 0.5--10~keV SIS band
(Table~2).
Nandra et al (1997a) have shown $\sigma^2_{rms}$ to be inversely 
correlated with X-ray luminosity for a sample of Seyfert 1 galaxies 
(including the 1993 data from NGC~3227).
As shown in Yaqoob et al (1997), the correlation between $\sigma^2_{rms}$
and X-ray luminosity remains when the 
lower value of $\sigma^2_{rms}$ found for the 1995 observation
of NGC~3227 is included.

\subsection{The Neutral-absorber}
\label{Sec:disc-neutral}

In all the models presented here we have assumed the observed 
spectrum is attenuated by screen of neutral material 
at $z=0$ (completely covering the cylinder--of--sight) with 
a column density $N_{H,0}^{gal} = 2.1\times10^{20}\ {\rm cm^{-2}}$.
This screen is adopted to model absorption within our galaxy,
with $N_{H,0}^{gal}$ fixed at the value 
derived from 21~cm measurements 
towards NGC~3227 (Murphy et al. 1996). 
However the analysis of the joint {\it ASCA}/{\it ROSAT} datasets 
from 1993 (\S\ref{Sec:mean-1993}) 
implies an additional screen of neutral material
($N_{H,z}^{neu} \simeq (4\pm1)\times10^{20}\ {\rm cm^{-2}}$)
which we assume is intrinsic to NGC~3227.
The 1995 observations are also consistent with the presence of 
such a screen.

From observations carried out in 1988 using a beam of
FWHM $\sim 13$~arcsec (i.e. probing a region 
$\sim3\times10^3$~light-years across within NGC~3227), 
Mundell et al (1995a) found evidence for H{\sc i} absorption at the 
systemic redshift of NGC~3227, implying an average column density 
$N_{H,z}^{HI} \sim 6\times10^{20}\ {\rm cm^{-2}}$
towards the nucleus.
More recent observations, taken with an angular resolution 
a factor $\sim 3$ higher,
imply $N_{H,z}^{HI} \simeq (11\pm1)\times10^{20}\ {\rm cm^{-2}}$ 
(Mundell, private communication).
Interestingly, Meixner et al (1990) detected elongated CO emission 
straddling the nucleus.
As noted by Mundell et al (1995a) this emission may represent the 
structure that is responsible for collimating the ultraviolet radiation
giving rise to the anisotropic \verb+[+O{\sc iii}\verb+]+ emission
observed by Mundell et al (1995b). 
Our {\it ASCA} observations imply that the column density of 
neutral material within the narrow cylinder--of--sight to the central 
X-ray source (with a diameter $\lesssim 10^4$~light-seconds) is 
a factor $\sim$2 smaller than the mean value of 
$N_{H,z}^{HI}$ on scale-sizes a factor $\sim 10^6$ larger.

We postpone the discussion of the large addition column density 
observed during the 1995 observations 
(which as noted in \S\ref{Sec:mean-1995} is formally consistent 
with neutral material) until \S\ref{Sec:disc-variable-abso}

\subsection{The Ionized-absorber}
\label{Sec:disc-ionized}

The 1993 data provide evidence for absorption by a screen of ionized 
material with column density 
$N_{H,z}^{ion} \simeq 3\times10^{21}\ {\rm cm^{-2}}$  and an 
X-ray ionization parameter $U_X \sim 0.01$. 
Absorption by ionized material is believed to be a common feature
in the X-ray spectra of Seyfert galaxies (Reynolds 1997; G98) and these values 
of $N_{H,z}^{ion}$ and $U_X$ are some of the lowest yet observed. 
G98 found a range 
$10^{21} \lesssim N_{H,z}^{ion} \lesssim 10^{23}\ {\rm cm^{-2}}$
and $0.01 \lesssim U_X \lesssim 0.1$ for ionized absorbers in 
a sample of Seyfert~1 galaxies. 
NGC~3227 has a relatively low X-ray luminosity 
($\sim 10^{42}\ {\rm erg\ s^{-1}}$ in the 2--10~keV band)
compared to other Seyfert~1s in that sample 
($\sim 10^{43}$--$10^{45}\ {\rm erg\ s^{-1}}$), and this may affect 
conditions in the absorber. 
Also, as noted in \S\ref{Sec:intro}, 
the optical depths of O{\sc vii} and O{\sc viii} 
are $\lesssim 0.05$ for $N_{H,z}^{ion} \lesssim 10^{21}\ {\rm cm^{-2}}$ 
making it difficult to detect with the signal-to-noise ratio of 
most {\it ASCA} observations. 

The column density appeared to increase between 1993 and 1995, 
consistent with a cloud moving into the cylinder--of--sight. 
As discussed in \S\ref{Sec:mean-1995}, 
the {\it ASCA}\ data are only able to constrain the ionization parameter to be 
$U_X \lesssim 0.06$ during this epoch (i.e. consistent with 
neutral material). However, the 1993 data  clearly show evidence 
that the absorber is ionized and in the context of this model 
the appearance of such a cloud 
increased the column of ionized material by an order of magnitude. 
These results are consistent with 
the suggestion of Turner \& Pounds (1989)
that the differential variability between the
soft and medium X-ray bands seen in the 
{\it EXOSAT}\ observations of NGC~3227 is most likely due to variations 
in the absorbing column. {\it Ginga}\ observations revealed evidence 
for an Fe $K$-shell absorption edge, with a depth corresponding 
to $N_{H,z}^{ion} \sim$few$\times10^{22}\ {\rm cm^{-2}}$
(Nandra \& Pounds 1994).
The current {\it ASCA}\ data do not allow any meaningful 
constraints to be placed on such a feature.

\subsubsection{A Dusty warm-absorber ?}
\label{Sec:dusty}

Dust is clearly present at some locations within the host galaxy,
as revealed by the detection of molecular emission and 
absorption features (see \S\ref{Sec:intro}) and implied by 
two thermal components observed in the 
10--300$\mu$m band (Rodr\'{i}guez Espinosa et al 1996).
In common with many other Seyfert galaxies, the component dominating 
the 10--30$\mu$m band (with a temperature $\sim$160~K) is likely to
arise as a result of the heating of dust by radiation sources 
$\lesssim 10$~kpc from the nucleus 
(e.g. Rodr\'{i}guez Espinosa \& P\'{e}rez Garc\'{i}a 1997).
Both the active nucleus and emission from any starburst
activity provide such radiation sources.
The question here is whether any dust exists within the 
immediate circumnuclear regions ($\lesssim 10$~pc) 
and, if so, whether it is in any way related to the ionized absorber.

As reviewed by K97, there are a number of indicators which can be used to 
probe the circumnuclear environment of NGC~3227. First, 
the underlying continuum in the optical and UV bands is unusually red 
compared to other Seyfert galaxies (Winge et al 1995). 
Using the ratio of the observed fluxes
at 125~nm and 220~nm (rest frame, which are relatively free of line emission)
and a standard Galactic composition for the dust, G98
found a column density $\sim 5\times10^{21}\ {\rm cm^{-2}}$
brought the intrinsic flux ratio into agreement with that 
observed in most other unobscured Seyfert galaxies.
Making slightly different assumptions K97 obtained a similar result.
Second, 
from the observed ratio of the broad H$\alpha$ and H$\beta$ lines
and assuming a Galactic dust/gas mass ratio, K97 
calculate an effective hydrogen column density to the broad emission line
region of $\sim 3\times10^{21}\ {\rm cm^{-2}}$.
As noted by Netzer (1990) the use of such line ratios in this way 
is potentially misleading as the intrinsic Balmer decrement is uncertain.
Nevertheless, we note that the implied column density is similar to 
that required to produce the observed reddening of the 
optical/UV continuum. Thus it seems likely that the same (dusty)
material is responsible for these observational characteristics.
The material must lie at a radius larger than of 
the BELR 
($\sim$10--20 light-days, Salamanca et al 1994; Winge et al 1995)
both in order to give rise to a depressed 
ratio of the intensity of broad H$\alpha$ and H$\beta$ lines, 
and since dust is unable to survive for long in 
the intense radiation field of the 
central source at smaller radii (e.g. Laor \& Draine 1993).
Interestingly the column density implied from such arguments 
($\sim$few$\times10^{21}\ {\rm cm^{-2}}$)
is very similar to that of the ionized absorber implied by the 
{\it ROSAT} observations (K97) and that 
derived from the {\it ASCA}\ observations in 1993
(above; G98).
It seems unlikely that this is purely coincidental.
As suggested by K97 and G98, it is more plausible that a substantial 
fraction of the ionized gas in NGC~3227 contains embedded dust, 
thus placing the ionized absorber {\it outside} the BELR.

Dusty ionized absorbers have also been suggested 
in the infrared-bright quasar IRAS~133349+2438 
(Brandt, Fabian \& Pounds 1996) and the Seyfert 
MCG--6-30-15 (Reynolds \& Fabian 1995; G98).
As pointed out in K97, the presence of dust can give rise to different model 
spectra, primarily as a result of the different depletion of the various 
elements in the gas-phase.
We have not attempted a detailed treatment of such cases here as most of 
the differences occur at energies below the {\it ASCA}\ bandpass.
If the dust is similar to that observed within the interstellar medium of 
the Galaxy, the composition of the gas-phase will be depleted in 
C and O by factor of $\sim 0.4$ and $\sim 0.6$ (respectively)
compared to "cosmic composition", and N and Ne suffering no significant 
depletion (e.g. Cowie \& Songaila 1986).
As can be from Fig.~\ref{fig:mean_ufmodelratio}, the main features 
imprinted on the underlying continuum in the {\it ASCA}\ bandpass 
by the ionized absorber are due to O{\sc vii} and O{\sc viii}. 
Thus if there is indeed embedded dust, and thus O is depleted 
in the gas-phase, one might expect 
the values of $N_{H,z}^{ion}$ derived above to have been underestimated 
by a factor $\sim 1.7$.
The differential depletion of the various elements gives rise 
to a subtle changes in the form of the spectrum across the 
0.6--1~keV band. Given the interplay between $N_{H,z}^{ion}$ and 
$U_X$ during the spectral fitting of {\it ASCA}\ data, the effect of 
including dust is more likely to increase the best-fitting value of 
$U_X$ (by a factor $\sim 3$ compared to dust-free models)
rather than $N_{H,z}^{ion}$.
Future observations with the grating-spectrometers 
onboard {\it AXAF} and {\it XMM} will allow a direct comparison 
between the depths of the edges from the highly-ionized, gas-phase ions 
and those of the neutral, dust-phase species. Such observations 
will therefore provide stringent constraints on any dust embedded 
within the ionized gas.

\subsubsection{Constraints for Photoionization Equilibrium}
\label{Sec:equilibrium}

In the absence of other sources of heating and cooling, the gas will be in
photoionization equilibrium if the photoionization and recombination
timescales of the dominant species 
($t_{ion}$ and $t_{rec}$ respectively)
are shorter than the timescale
($t_{var}$) for large-amplitude variations in the intensity of 
the illuminating continuum.
To first order $t_{ion} \simeq N_{H,z} /U_X n c\ {\rm s}$.
Thus substituting the mean values derived from the 1993 observations
($N_{H,z}^{ion} = 3\times10^{21}\ {\rm cm^{-2}}$ and $U_X = 0.01$), we 
have $t_{ion} \simeq 10^{13} n^{-1}\ {\rm s}$.
The recombination time of O{\sc viii} is longer than that for O{\sc vii}
and given by 
$t_{rec}$(O{\sc viii}) $\simeq 3\times10^{8} T^{0.5} n^{-1}\ {\rm s}$
for a gas at an equilibrium temperature $T$
(e.g. Verner \& Ferland 1996).
Thus for plausible values of $T$ ($\sim 10^5$~K), 
$t_{ion}$ is more important than $t_{rec}$(O{\sc viii})
when determining whether or not material illuminated by 
a variable ionizing continuum will be in equilibrium.
Thus the material will be equilibrium ($t_{var} > t_{ion}$)
when $ n \gtrsim 10^{13} t_{var}^{-1}\ {\rm cm^{-3}}$.
Unfortunately the assignment of an appropriate value for 
$t_{var}$ is problematic since little is know about the 
detailed variability characteristics of NGC~3227 in the X-ray band
and, as a class, Seyfert 1 galaxies appear to exhibit 
a 'red-noise' power-density spectrum
(e.g. Lawrence \& Papadakis 1993; Green, McHardy \& Lehto 1993, and 
references therein).
However for illustration, here we assume 
$t_{var} = 2\times10^{4}\ {\rm s}$ (the approximate
timescale in which factor $\sim$2 changes in flux are observed
during the temporal period 1993(t3) 
in Fig.~\ref{fig:lc128}), leading to the requirement that
$ n \gtrsim 5\times10^{8}\ {\rm cm^{-3}}$ for the ionized absorber 
to be in equilibrium.
Such a density is consistent with 
that assumed in our {\tt ION} models for the ionized material.
From the definition of $U_X$ and assuming 
$L_{0.1-10} = 1.6\times10^{42}\ {\rm erg\ s^{-1}}$, such 
values of $n$ require $r \lesssim 10$ light-days. 
Thus, if the ionized absorber contains embedded dust
(as suggested in \S\ref{Sec:dusty}) it must be at a larger 
radius and hence will {\it not} be in perfect equilibrium
for some time after the temporal period 1993(t3).
Unfortunately, however, it is impossible to quantify  
just how far out of equilibrium the ionized material 
may have become, or the time taken 
for equilibrium to be restored, without a detailed 
knowledge of the ionized material, 
and a complete knowledge of the variations in the ionizing 
continuum both before temporal period 1993(t1) and after 
temporal period 1993(t4).

\subsection{The Variable Absorption}
\label{Sec:disc-variable-abso}

In \S\ref{Sec:mean-cont} we found the absorbing column density to 
have increased by a factor $\sim10$
(from $N_{H,z}^{ion} \sim 3\times10^{21}\ {\rm cm^{-2}}$ 
to $\sim 30\times10^{21}\ {\rm cm^{-2}}$)
between the 1993 and 1995 observations.
As noted in \S\ref{Sec:mean-1995} the {\it ASCA}\ data
do not allow us to place stringent constraints on the ionization state 
of the material 
during the latter epoch, with value of $U_X=0$ (neutral material) 
to $U_X \sim 0.06$ (highly ionized) being consistent with the 
observations.
We suggest that the most likely explanation is that a cloud of material 
moved into the cylinder--of--sight. 
The location and kinematics of such a cloud clearly cannot be determined with
the current data.
However within some radius Keplarian motion (alone) is able
to move material completely through the cylinder--of--sight (of diameter equal to
the size of the X-ray emitting region) in the 2 years between the observations.
The X-ray emitting region in NGC~3227 can be estimated 
to be $10^{4}$ light-seconds across from the variability behaviour 
observed during 1993. 
From G98, the fiducial radius is 
$r_{\rm ld} \lesssim 20 
(f_{\rm bolX}/ f_{\rm Edd}) L_{X42}$ light-days, 
where $f_{\rm bolX}L_{X42}\times10^{42}\ {\rm erg\ s^{-1}}$ 
is the bolometric luminosity and 
$f_{\rm Edd}$ is the fraction of the Eddington luminosity 
at which the object is emitting.
Thus, for NGC~3227 (with $L_{X42} \sim 1$)
and assuming $f_{\rm bolX}/ f_{\rm Edd} \sim $~few, it is 
feasible that tranverse motion of a cloud within the 
inner regions of the AGN responsible for the 
change in column density.
It would be interesting to see whether any of the 
other reddening indicators in NGC~3227 vary on a timescale of
years, especially the reddening towards the BELR.

\subsection{The Unattenuated Component}
\label{Sec:disc-unattenuated}

During the 1995 observations we found spectroscopic evidence for 
$\sim 13$\% of the observed continuum to be unattenuated
by the absorber with $N_{H,z}^{ion} \simeq 30\times10^{21}\ {\rm cm^{-2}}$.
Such a model is consistent with 
both the case where only a fraction of the cylinder--of--sight to NGC~3227 
is covered by the absorbing material (say in the form of clouds), and to the 
case where the whole cylinder--of--sight is covered by the material 
but in which a fraction of the continuum escapes by another 
light-path.
We consider both explanations equally plausible. 
The latter, however, raises the question as to the location and conditions 
within the material responsible for scattering the unattenuated component 
back into our cylinder--of--sight.
Since we find no evidence for any strong emission or absorption features 
in the spectrum at energies 
$\lesssim 1$~keV, the material must be highly ionized.
In particular we see no evidence for any emission lines due to the
H-like and He-like species of the abundant elements which are 
commonly seen in the scattered X-ray spectra of Seyfert~2 galaxies
(e.g. Turner et al 1997), with upper limits $\sim 30$~eV
at 90\% confidence.
Unfortunately any variations in the XM$_1$ and XM$_3$ bands 
during the 1995 observations
(which are dominated by the scattered and 'transmitted' components 
respectively under this hypothesis) are of insufficient amplitude 
to offer any insight into the location of the material responsible 
for scattering.

Finally, we note that the X-ray spectrum of NGC~3227 during this epoch is 
rather similar to that of NGC~4151 (Weaver et al 1994a,b; G98), 
although the latter source has $D_f \sim 5$\%
and a somewhat larger value of $N_{H,z}^{ion}$ at most epochs.

\subsection{The Fe Regime}
\label{Sec:disc-line}

Our data confirm the presence of Fe $K$-shell emission within NGC~3227,
with an equivalent width $\sim$100--300~eV.
We find evidence that the emission line is broad, most likely with an 
asymmetric profile.
Thus our findings confirm the suggestion from the earlier 
{\it Ginga}\ observations (e.g. George, Nandra \& Fabian 1990), and 
show that the Fe $K$-shell emission within NGC~3227 is 
fairly typical 
of other Seyfert 1 galaxies (e.g. Nandra et al
1997b, and references within)
and some Seyfert 2 galaxies (Turner et al 1998).
We find no compelling evidence that the Fe emission varied in either
shape or equivalent width either within the individual observations 
or between the 1993 and 1995 epochs. Thus we find no evidence for 
significant changes in the distribution of Fe emissivity.

\section{CONCLUSIONS}
\label{Sec:Conc}

{\it ASCA}\ observations of NGC~3227
performed during 1993 and 1995 show 
evidence for marked spectral variability both within 
and between the observations. The source 
shows evidence for a column 
$N_{H,z}^{ion} \simeq 3\times10^{21}\ {\rm cm^{-2}}$ 
of ionized material with an X-ray ionization parameter 
$U_X \simeq 0.01$ during 1993, increasing by an 
order of magnitude by the 1995 epoch.  
The slope of the continuum steepened 
by $\Delta \Gamma \simeq 0.1$ during a flare 
within the 1993 observation. However, the data do not allow us to 
distinguish between a steepening of the 
'primary' continuum, or a change in the relative strengths of the power-law 
and a putative Compton-reflection component.

\acknowledgements

We thank Lorella Angelini (NASA/GSFC) for useful 
discussions, and Carole Mundell (Univ. Maryland)
for communication of her results 
prior to publication.
We acknowledge the financial support of the 
Universities Space Research Association (IMG, TJT, TY),
NASA (AP), 
the National Research Council (KN), and
a special grant from the Israel Science Foundation \& 
by the Jack Adler Chair of Extragalactic Astronomy (NH). 
This research has made use of 
the Simbad database, operated at CDS, Strasbourg, France; and
of data obtained through the HEASARC on-line service, provided by
NASA/GSFC.

\clearpage

\clearpage
\typeout{FIGS}

\clearpage

\clearpage

\vspace{10cm}
{\it Caption for Fig. 3.}\\
The results from the analysis of the time-averaged spectra 
of the 1993 (plots a,b \& c) 
and 1995 (plots d \& e) {\it ASCA}\ observations of NGC~3227,
excluding the Fe $K$-shell band (see \S\ref{Sec:mean-cont}). 
In each
case the upper panel shows (in bold) the best-fitting model, 
the model after correcting for all neutral absorption 
($N_{HI}^{Gal}$ and $N_{H,z}^{neu}$), and (dashed) the implied underlying continuum. 
The lower panel shows the mean data/model ratios. The filled triangles show the
(error-weighted) means of the ratios from the individual {\it ASCA}\ detectors
for the energy-bands used in the spectral analysis, rebinned in energy-space
for clarity. (Note the different y-axis scale in plot a.)
The dotted errorbars show the corresponding rebinned, mean ratios
when the best-fitting model is extrapolated $<0.6$~keV and into the 5--7~keV
band. 
The stars show data/model ratios of the 1993 {\it ROSAT}\ PSPC data, 
again
rebinned in energy-space for clarity. The details of the individual panels are
described in \S\ref{Sec:mean-cont}.

\clearpage
\thispagestyle{empty}
\clearpage

\clearpage

\thispagestyle{empty}
\clearpage

\thispagestyle{empty}
\clearpage

\clearpage

\typeout{END FIGS}


\clearpage

\begin{references}
\reference{Ar66} Arp, H., 
		1966, {\it Atlas of Peculiar Galaxies}, 
		(CalTech: Pasadena), plate 94
\reference{Ar94} Arribas, S., Mediavilla, E.,
		1994, apj, 437, 149
\reference{Ba93} Baan, W.A., Haschick, A.D., Uglesich, R., 
		1993, \apj, 415, 140
\reference{Br96} Brandt, W.N., Fabian, A.C., Pounds, K.A., 
		1996, \mnras, 278, 326
\reference{Br94} Briel, U., et al 
		1994, {\it The ROSAT User's Handbook}
\reference{Co86} Cowie, L.L., Songaila, A., 
		1986, ARAA, 24, 499
\reference{Cr97} Crenshaw, D.M.,
                1997, In {\it Emission Lines in Active Galaxies:
                New Methods and Techniques},
                eds. Peterson, B.M, Cheng, F.-Z., Wilson, A.S.
                (ASP, San Fransisco), p.240
\reference{Cr98} Crenshaw, D.M., Maran, S.P., Mushotzky, R.F.,
		1998, \apj, 496, 797
\reference{Fa89} Fabian, A.C., Rees, M.J., Stella, L., White, N.E.,
		1989, \mnras, 238, 729
\reference{Fa94} Fabian, A.C., et al,
		1994, \pasj, 46, L59
\reference{FGF}  George, I.M., Fabian, A.C., 
		1991, \mnras, 249, 352
\reference{Ge90} George, I.M., Nandra, K., Fabian, A.C.,
		1990, \mnras, 242, 28P
\reference{Ge98a} George, I.M., Netzer, H., Turner, T.J., Nandra, K., 
		Mushotzky, R.F., Yaqoob, T.,
	 	1998a, \apjs, 114, 73 (G98)
\reference{Ge98b} George, I.M., Turner, T.J., Mushotzky, R.F., Nandra, K., 
		Netzer, H., 
	 	1998b, \apj, in press 
\reference{GH94} Ghisellini, G., Haardt, F., 
		1994, \apj, 429, L53
\reference{Go97} Gonz\'{a}lez Delgado, R.M., P\'{e}rez, E.,
		 1997, \mnras, 284, 931
\reference{Gr93} Green, A.R., McHardy, I.M., Lehto, H.J., 
		1993, \mnras, 265, 664
\reference{Gu96} Guainazzi, M., Mihara, T., Otani, C., Matsuoka, M.,
		1996, \pasj, 48, 781
\reference{Haa97} Haardt, F., Maraschi, L., Ghisellini, G., 
		1997, \apj, 476, 620
\reference{Ha84} Halpern, J.P.,
		1984, \apj, 281, 90
\reference{Hu92} Huchra, J.P., Burg, R., 
		1992, \apj, 393, 90
\reference{Jo93} Johnson, W.N., et al.,
		1993, \aaps, 97, 21
\reference{Ki76} Kikoin, I.K., (ed)
		1976, {\it Tables of Physical Quantities},
		Atomizdat, Moscow.
\reference{Ko97} Komossa, S., Fink, H., 
		1997, \aap, 327, 483 (K97)
\reference{Kr96} Kriss, G.A., et al., 
		1996, \apj, 467, 629
\reference{Ari93} Laor, Ari, Draine, B.T., 
		1993, \apj, 402, 441
\reference{La93} Lawrence, A., Papadakis, I., 
		1993, \apj, 414, L85
\reference{Mak96} Makishima, et al 
		1996, \pasj, 48, 171 
\reference{Ma97} Malizia, A., Bassani, L., Stephen, J.B., 
		Malaguti, G., Palumbo, G.G.C., 
		1997, \apjs, 113, 311
\reference{Smita} Mathur, S., 
		1994, \apj, 431, L75    
\reference{Me93} Mediavilla, E., Arribas, S., 
		1993, \nat, 365, 420
\reference{Me90} Meixner, M., Puchalsky, R., Blitz, L., Wright, M., Heckman, T.,
		1990, \apj, 354, 158
\reference{Mu95a} Mundell, C.G., Pedlar, A., Axon, D.J., Meaburn, J., 
			Unger, S.W.,
		1995a, \mnras, 277, 641
\reference{Mu95b} Mundell, C.G., Holloway, A.J., Pedlar, A., Meaburn, J., 
		Kukula, M.J., Axon, D.J., 
		1995b, \mnras, 275, 67
\reference{Mur96} Murphy, E.M., Lockman, F.J., Laor, A., Elvis, M., 
		1996, \apjs, 105, 369
\reference{NP94} Nandra, K., Pounds, K.A.,
		1994, \mnras, 268, 405
\reference{Na97a} Nandra, K., George, I.M., Mushotzky, R.F., Turner, T.J.,
                 Yaqoob, T.,
                   1997a, \apj, 476, 70
\reference{Na97b} Nandra, K., George, I.M., Mushotzky, R.F., Turner, T.J.,
                 Yaqoob, T.,
                   1997b, \apj, 477, 602
\reference{Ne90} Netzer, H., 
		1990, in {\it Active Galactic Nuclei}, ed. R.D. Blandford,
		H.Netzer, L.Woltjer (Berlin: Springer), p107
\reference{Ne93} Netzer, H.,
                1993, \apj, 411, 594
\reference{Ne96} Netzer, H.,
                1996, \apj, 473, 781
\reference{Ne94} Netzer, H., Turner, T.J., George, I.M.,
		1994, \apj, 435, 106
\reference{Or98} Orr, A., Yaqoob, T., Parmar, A.N., Piro, L., White, N.E., 
		Grandi, P., 
		1998, \aap, submitted
\reference{Os93} Osterbrock, D.E., Martel, A., 
		1993, \apj, 414, 552
\reference{Ot96} Otani, C., et al.,
		1996, \pasj, 48, 211
\reference{Po89} Pounds, K.A., Nandra, K., Stewart, G.C, Leighly, K.,
		1989, \mnras, 240, 769
\reference{Pt94} Ptak, A., Yaqoob, T., Serlemitsos, P.J., Mushotzky, R.F.,
		Otani, C.,
		1994, \apj, 436, L31
\reference{Pt98} Ptak, A., et al.,
		1998, in preparation
\reference{Ra97} Radecke, H.-D., 
		1997, \aap, 319, 18
\reference{Re97} Reynolds, C.S., 
		1997, \mnras, 286, 513
\reference{RF95} Reynolds, C.S., Fabian, A.C.,
		1995, \mnras, 273, 1167
\reference{Re95} Reynolds, C.S., Fabian, A.C., Nandra, K., Inoue, H., 
		Kunieda, H., Iwasawa, K., 
		1995, \mnras, 277, 901
\reference{Ri82} Rickard, L.J., Bania, T.M., Turner, B.E., 
		1992, \apj, 252, 147
\reference{Ri97} Rigopoulou, D., Papadakis, I., Lawrence, A., Ward, M.,
		1997, \aap, 327, 493
\reference{Ro96} Rodr\'{i}guez Espinosa, J.M., P\'{e}rez Garc\'{i}a, A.M., 
		Lemke, D., Meisenheimer, K., 
		1996, \aap, 315, L129.
\reference{Ro97} Rodr\'{i}guez Espinosa, J.M., P\'{e}rez Garc\'{i}a, A.M., 
		1997, \apj, 487, L33.
\reference{Sa94} Salamanca, I., et al,
		1994, \aap, 282, 742
\reference{Sh97} Shields, J.C., Hamann, F., 
		1997, \apj, 481, 752
\reference{Sn95} Snowden, S.L, Turner, T.J., George, I.M, Yusaf, R., 
		Predehl, P., Prieto, A.,
		1995, {\it OGIP Calibration Memo}  CAL/ROS/95-003
\reference{St95} Stern, B., Poutanen, J., Svensson, R., Sikora, M., 
		Begelman, M.C., 
		1995, \apj, 449, 13
\reference{Su90} Sunyaev, R.A., Titarchunk, L.G., 
		1980, \aap, 86, 121
\reference{Sv94} Svensson, R., 
		1994, \apjs, 92, 585
\reference{Sv96} Svensson, R., 
		1996, \apjs, 120, 475
\reference{TP89} Turner, T.J., Pounds, K.A.,
		1989, \mnras, 240, 833
\reference{Tu91} Turner, T.J., Weaver, K.A., Mushotzky, R.F., Holt, S.S.,
		Madejski, G.M., 
		1991, \apj, 381, 85
\reference{Tu97} Turner, T.J., George, I.M., Nandra, K., Mushotzky, R.F.,
		1997, \apjs, 113, 23
\reference{Tu98} Turner, T.J., George, I.M., Nandra, K., Mushotzky, R.F.,
		1998, \apj, 493, 91
\reference{V96} Verner, D.A., Ferland, G.J., 
		1996, \apjs, 103, 467
\reference{We95} Weaver, K.A., Arnaud, K.A., Mushotzky, R.F., 
		1995, \apj, 447, 121
\reference{We94a} Weaver, K.A., et al., 
		1994a, \apj, 423, 621
\reference{We94b} Weaver, K.A., Yaqoob, T., Holt, S.S., Mushotzky, R.F., 
		Matsuoka, M., Yamauchi, M., 
		1994b, \apj, 436, L27
\reference{Wi95} Winge, C., Peterson, B.M., Horne, K., Pogge, R.W.
		Pastoriza, M.G., Storchi-Bergmann, T., 
		1995, \apj, 445, 680
\reference{Ya97} Yaqoob, T., Nandra, K., George, I.M., Turner, T.J.,
		1997, in {\it All-sky X-ray Observations in the 
		Next Decade}, ed. M.Matsuoka, N.Kawai
		(Wako, Japan: RIKEN), 219 
\end{references}
\end{document}